\documentclass[conference]{IEEEtran}

\usepackage[noadjust]{cite}
\usepackage{amsmath,amssymb,amsfonts}
\usepackage{mathtools}
\usepackage[hyphens]{url}

\usepackage{wasysym}

\usepackage{algorithmic}
\usepackage{graphicx}
\usepackage{textcomp}
\usepackage{xcolor}
\def\BibTeX{{\rm B\kern-.05em{\sc i\kern-.025em b}\kern-.08em
    T\kern-.1667em\lower.7ex\hbox{E}\kern-.125emX}}

\usepackage[utf8]{inputenc}
\usepackage[english]{babel}

\usepackage{bookmark}
\usepackage{varioref} %
\usepackage{hyperref} %
\usepackage{cleveref} %
\crefname{lstlisting}{listing}{listings}
\Crefname{lstlisting}{Listing}{Listings}

\usepackage{tabularx}
\usepackage{booktabs}
\usepackage{placeins}
\usepackage{caption}
\usepackage{subcaption}
\usepackage{comment}
\usepackage{enumerate}
\usepackage{xspace}

\usepackage{multirow}

\usepackage{listings}
\usepackage{flushend}

\usepackage{enumitem}
\setlist[itemize]{noitemsep, topsep=0pt, leftmargin=10pt}

\usepackage{colortbl}
\definecolor{Gray}{gray}{0.65}
\definecolor{LightGray}{gray}{0.9}

\newcommand{\fakeparagraph}[1]{\vskip 0pt\noindent\textbf{#1 }}

\usepackage{bold-extra}
\usepackage[printonlyused,nohyperlinks]{acronym}

\usepackage[binary-units,group-separator={,},range-units=single,range-phrase=--]{siunitx}
\DeclareSIUnit[number-unit-product = {\thinspace}]{\inch}{inch}
\DeclareSIUnit{\bits}{bits}
\DeclareSIUnit{\bit}{bit}
\DeclareSIUnit \px {\ensuremath{\mathit{px}}}

\usepackage{booktabs}

\usepackage{multirow}  	 	

\usepackage{tabularx,ragged2e} 	

\usepackage{mathtools}  
\DeclarePairedDelimiter\nearest{\lfloor}{\rceil}

\DeclarePairedDelimiterX{\reducedX}[2]{[}{]_{#1}}{#2}

\newcommand{\cpp}{C\texttt{++}\xspace}

\newcommand{\AndGate}{AND\xspace}
\newcommand{\Xor}{XOR\xspace}
\newcommand{\OrGate}{OR\xspace}
\newcommand{\NorGate}{NOR\xspace}
\newcommand{\MuxGate}{MUX\xspace}

\newcommand{\Ramparts}{\textsc{Ramparts}\xspace}
\newcommand{\Palisade}{PALISADE\xspace}

\newcommand{\Lol}{\texorpdfstring{$\Lambda\!\circ\!\lambda$}{Lol}\xspace}

\newcommand{\fhew}{FHEW\xspace}

\newcommand{\pk}{\texttt{pk}}
\newcommand{\sk}{\texttt{sk}}

\newcommand{\R}{\mathbb{R}}
\newcommand{\Z}{\mathbb{Z}}

\newcommand{\T}{\mathbb{T}}
\newcommand{\C}{\mathbb{C}}

\usepackage{adjustbox}
\usepackage{array}
\newcolumntype{R}[2]{%
    >{\adjustbox{angle=#1,lap=\width-(#2)}\bgroup}%
    l%
    <{\egroup}%
}
\newcommand*\rot{\multicolumn{1}{R{45}{1em}}}

\DeclareMathOperator{\Enc}{Enc}
\DeclareMathOperator{\Dec}{Dec}

\begin{document}

\newacro{alchemy}[\textsc{Alchemy}]{A Language and Compiler for Homomorphic Encryption Made easY}
\newacro{ast}[AST]{Abstract Syntax Tree}
\newacro{aes}[AES]{Advanced Encryption Standard}

\newacro{bfv}[BFV]{Brakerski/Fan-Vercauteren}
\newacro{bgv}[BGV]{Brakerski-Gentry-Vaikuntanathan}
\newacro{bfs}[BFS]{Breadth-First Search}

\newacro{cpu}[CPU]{Central Processing Unit}
\newacro{cggi}[CGGI]{Chillotti-Gama-Georgieva-Izabachene}
\newacro{cgt}[CGT]{Class Generation Tool}
\newacro{ckks}[CKKS]{Cheon-Kim-Kim-Song}
\newacro{crt}[CRT]{Chinese Remainder Theorem}
\newacro{cfg}[CFG]{Control Flow Graph}
\newacro{ctr}[CTR]{Click-Through Rate}
\newacro{ccpa}[CCPA]{California Consumer Protection Act}
\newacro{chet}[CHET]{Compiler and Runtime for Homomorphic Evaluation of Tensor Programs}
\newacro{cnn}[CNN]{Convolutional Neural Network}
\newacro{cmux}[CMux]{Conditional Multiplexer}
\newacro{ci}[CI]{Continuous Integration}

\newacro{dsl}[DSL]{Domain-Specific Language}
\newacro{dft}[DFT]{Discrete Fourier Transform}
\newacro{dag}[DAG]{Directed Acyclic Graph}
\newacro{dfs}[DFS]{Depth-First Search}
\newacro{dfg}[DFG]{Data Flow Graph}
\newacro{ml}[ML]{Machine Learning}

\newacro{e3}[E\textsuperscript{3}]{Encrypt-Everything-Everywhere}
\newacro{e2ee}[E2EE]{End-to-End Encryption}
\newacro{eva}[EVA]{Encrypted Vector Arithmetics Language and Compiler}

\newacro{fft}[FFT]{Fast Fourier Transformation}
\newacro{fhe}[FHE]{Fully Homomorphic Encryption}
\newacro{ftt}[FTT]{Fermat-Theoretic Transform}
\newacro{flash}[FLaSH]{Fully, Leveled and Somewhat Homomorphic Encryption Library}

\newacro{gsw}[GSW]{Gentry-Sahai-Waters}
\newacro{gpv}[GPV]{Gentry-Peikert-Vaikuntanathan}
\newacro{gpu}[GPU]{Graphics Processing Unit}
\newacro{gwas}[GWAS]{Genome-Wide Association Studies}
\newacro{gdpr}[GDPR]{General Data Protection Regulation}

\newacro{he}[HE]{Homomorpic Encryption}
\newacro{helib}[HElib]{Homomorphic Encryption Library}
\newacro{heaan}[HEAAN]{Homomorphic Encryption for Arithmetic of Approximate Numbers}

\newacro{ir}[IR]{Intermediate Representation}
\newacro{ide}[IDE]{Integrated Development Environment}

\newacro{json}[JSON]{JavaScript Object Notation}

\newacro{lwe}[LWE]{Learning With Errors}
\newacro{lut}[LUT]{Look-Up Table}

\newacro{mpc}[MPC]{Multi-Party Computation}

\newacro{ntt}[NTT]{Number-Theoretic Transform}

\newacro{pahe}[PAHE]{Packed Additively Homomorphic Encryption}
\newacro{prf}[PRF]{Pseudorandom Function}

\newacro{rtl}[RTL]{Register-Transfer-Level}
\newacro{rns}[RNS]{Residue Number System}
\newacro{rlwe}[RLWE]{Ring-Learning With Errors}

\newacro{s-expression}[s-expression]{Symbolic Expression}
\newacro{seal}[SEAL]{Simple Encrypted Arithmetic Library}
\newacro{simd}[SIMD]{Single Instruction, Multiple Data}
\newacro{stst}[StSt]{Stehle-Steinfeld}
\newacro{she}[SHE]{Somewhat Homomorphic Encryption}
\newacro{stl}[STL]{Standard Template Library}

\newacro{tfhe}[TFHE]{Fast Fully Homomorphic Encryption Library over the Torus}

\newacro{uml}[UML]{Unified Modeling Language}

\newacro{wfa}[WFA]{Weighted Finite Automata}

\title{SoK: Fully Homomorphic Encryption Compilers}

\author{
\IEEEauthorblockN{Alexander Viand}
\IEEEauthorblockA{\textit{ETH Zurich}\\ %
alexander.viand@inf.ethz.ch}
\and
\IEEEauthorblockN{Patrick Jattke}
\IEEEauthorblockA{\textit{ETH Zurich}\\ %
pjattke@ethz.ch}
\and
\IEEEauthorblockN{Anwar Hithnawi}
\IEEEauthorblockA{\textit{ETH Zurich}\\ %
anwar.hithnawi@inf.ethz.ch}
}

\maketitle

\thispagestyle{plain}
\pagestyle{plain}

\begin{abstract}

Fully Homomorphic Encryption (FHE) allows a third party to perform arbitrary computations on encrypted data, learning neither the inputs nor the computation results. Hence, it provides resilience in situations where computations are carried out by an untrusted or potentially compromised party. This powerful concept was first conceived by Rivest et al. in the 1970s. However, it remained unrealized until Craig Gentry presented the first feasible FHE scheme in 2009.

The advent of the massive collection of sensitive data in cloud services, coupled with a plague of data breaches, moved highly regulated businesses to increasingly demand confidential and secure computing solutions. This demand, in turn, has led to a recent surge in the development of FHE tools. To understand the landscape of recent FHE tool developments, we conduct an extensive survey and experimental evaluation to explore the current state of the art and identify areas for future development.

In this paper, we survey, evaluate, and systematize FHE tools and compilers. We perform experiments to evaluate these tools' performance and usability aspects on a variety of applications.
We conclude with recommendations for developers intending to develop FHE-based applications and a discussion on future directions for FHE tools development.

\end{abstract}

\section{Introduction}
\label{intro}

Recent years have seen unprecedented growth in the adoption of cloud computing services. 
More and more highly regulated businesses and organizations (e.g., banks, governments, insurances, health), where data security is paramount, move their data and services to the cloud. 
This trend has led to a surge in demand for secure and confidential computing solutions that protect data confidentiality while in transit, rest, and in-use. 
This is an amply justified and expected demand, particularly in the light of the numerous reports of data breaches~\cite{saleem2020sok, enterprise-data-breach}. 
\acf{fhe} is a key technological enabler for secure computation and has recently matured to be practical for real-world use~\cite{Zama, Enveil2020-ds, Inpher2018-ry, Lunden2019-nx, Jain_undated-dh, Loritz_undated-aq,Osborne2020-yj}.

\ac{fhe} allows arbitrary computations to be performed over encrypted data, eliminating the need to decrypt the data and expose it to potential risk while in use.
While first proposed in the 1970s~\cite{Rivest1978a}, FHE was long considered impossible or impractical.
However, thanks to advances in the underlying theory, general hardware improvements, and more efficient implementations, it has become increasingly practical.
In 2009, breakthrough work from Craig Gentry proposed the first feasible FHE scheme~\cite{Gentry2009-zi}.
In the last decade, FHE has gone from a theoretical concept to reality,  %
with performance improving by up to five orders of magnitude.
For example, times for a multiplication between ciphertexts dropped from 30 minutes to less than 20 milliseconds.  
While this is still around seven orders of magnitude slower than an \texttt{IMUL} instruction on a modern CPU, it is sufficient to make many applications practical.
Additionally, modern schemes introduced SIMD-style parallelism, encoding thousands of plaintext values into a single ciphertext to further improve throughput~\cite{Smart2014}.

These advances have enabled a wide range of applications covering a wide range of domains.
These include mobile applications,
where FHE has been used to encrypt the back end of a privacy-preserving fitness app~\cite{Microsoft2019-qj},  while continuing to provide a real-time experience. 
In the medical domain, FHE has been used to enable privacy-preserving genome analysis~\cite{Kim2020-dk} applications over large datasets.
More generally, FHE has been used to solve various well-known problems like Private Set Intersection (PSI)~\cite{Chen2018a}, outperforming previous solutions by 2$\times$ in running time. %
In the domain of machine learning, FHE has been used for tasks ranging from linear and logistic regression~\cite{Kim2019} %
 to Encrypted Neural Network inference~\cite{Dathathri2019-vu}, which can be used to run privacy-preserving ML-as-a-Service applications, %
for example, for private phishing email detection~\cite{chou2020}.
 As a consequence, there has been increasing interest in FHE-based secure computation solutions~\cite{Zama, Enveil2020-ds, Inpher2018-ry, Lunden2019-nx, Jain_undated-dh, Loritz_undated-aq,Osborne2020-yj}.
 Gartner projects~\cite{Driver2020-gc} that ``by 2025, at least 20\% of companies will have a budget for projects that include fully homomorphic encryption."

    Despite these recent breakthroughs, building secure and efficient FHE-based applications remains a challenging task.
    This is largely attributed to the differences between traditional programming paradigms and \ac{fhe}'s computation model, which poses unique challenges.
    For example,
    virtually all standard programming paradigms rely on data-dependent branching, e.g., if/else statements and loops.
 On the other hand, FHE computations are, by definition, data-independent, or they would violate the privacy guarantees.    %
    Working with FHE also introduces significant engineering challenges in practice.
    Different schemes offer varying performance tradeoffs, and optimal choices are heavily application-dependent.
    To address some of the engineering challenges in this space, we have seen a surge of work on tools that aim to
    improve accessibility and reduce barriers to entry in this field.

  Without tool support, realizing  FHE-based computations by implementing the required mathematical operations directly or using an arbitrary-precision arithmetic library is complex, requiring considerable expertise in both cryptography and high-performance numerical computation.
    Therefore, \ac{fhe} libraries like the \acf{seal}~\cite{sealcrypto} or the \acf{tfhe}~\cite{TFHE} implement the underlying cryptographic operations and expose a higher-level API.
    In addition to key generation, encryption, and decryption, these APIs also expose at least homomorphic addition and multiplication.
    
    In practice, however, library APIs often include dozens of additional functionalities for ciphertext maintenance and manipulation.
    Since schemes vary in features, these APIs differ significantly not just in their implementation but also conceptually.
    Efforts are being made to standardize APIs for FHE schemes~\cite{HomomorphicEncryptionSecurityStandard} 
    and, simultaneously, there are first steps towards interoperability via wrappers around existing libraries~\cite{Barlow2019-si}.
    However, achieving competitive performance frequently still requires working with libraries directly.
    
\begin{figure}
        \centering
        \includegraphics[width=\linewidth]{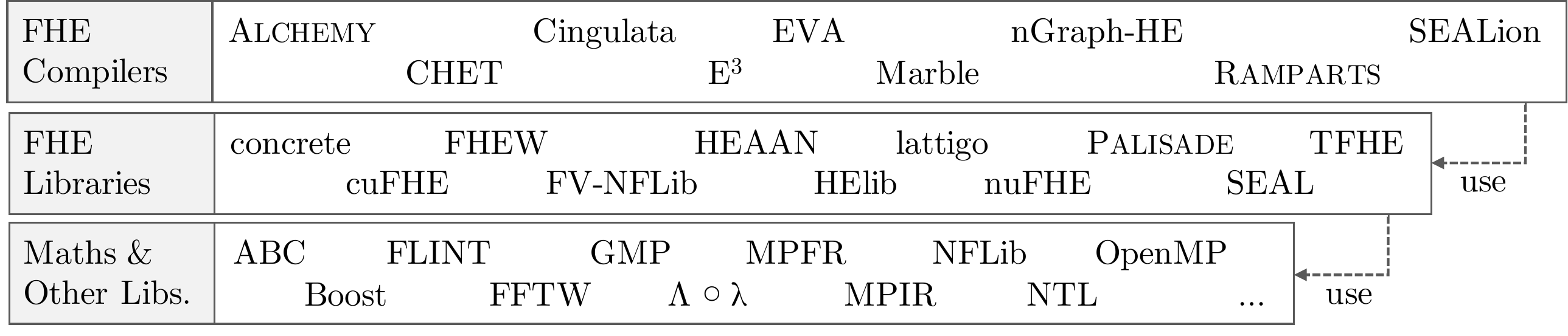}
        \caption{Overview of the FHE tool space.} %
        \label{fig:libs_pyramide}
    \end{figure}

    While FHE libraries make the process of writing FHE-based applications substantially more efficient, they still require significant expertise and understanding of the underlying scheme
     since they remain relatively low-level cryptographic libraries.
    Therefore, recent years have seen the development of higher-level tools, frequently known as FHE \emph{compilers}, that aim to translate standard programs into FHE-based implementations. 
 These tools focus on making FHE accessible to non-experts by improving usability and increasingly offering advanced optimizations previously accessible only to experts.
    Compilers generally rely on FHE libraries to realize the actual en-/decryption and homomorphic computation.
    FHE libraries, in turn, frequently employ existing libraries for fast numerical computations, parallelization, or other non-FHE-specific features.
    \Cref{fig:libs_pyramide} depicts different FHE tools and where they fit into this dependency hierarchy.

While much work remains to be done, these tools have significantly eased the task of developing FHE applications.
    For example, in the domain of machine learning, tools have demonstrated accessibility and usability while also
    providing state-of-the-art performance due to automatic optimizations that significantly outperform previous
    hand-crafted solutions by experts.
    The nGraph-HE framework~\cite{Boemer2019-mt}, for example, converts neural networks into efficient FHE-based implementations for private inference.
    Here, nearly all aspects related to FHE are abstracted away, and the user experience is essentially identical to working directly with TensorFlow.

    Though there has been a surge of works on FHE tools and accessibility, 
    we currently lack a comprehensive overview of the current state of FHE development.
    While it is clear that both significant advances have been made and many challenges remain open,
    there is no systematic understanding of the remaining engineering challenges that need to be addressed to help broaden FHE adoption.

    Therefore, this paper aims to fill in this knowledge gap by studying and surveying the current state-of-the-art of FHE tools.
    More concretely, this survey has two objectives: First, to assist developers looking to develop FHE-based applications in selecting a suitable approach and,
    second, to provide the community with valuable insights on both successes and remaining issues in this space.

    Towards this goal, we conduct an extensive survey of existing tools and highlight their features and characteristics.
    Subsequently, we consider these tools in practice by experimentally evaluating them across a range of case study applications, 
    contrasting usability, expressiveness, and performance. 

    In our experimental evaluation, we consider a selection of tools in more detail and provide an in-depth analysis of their usability and expressiveness in practice.
    We implement and benchmark three case-study applications that represent different domains of FHE-based computation.
    Our benchmarks allow us to study not only the overall performance of FHE for these applications across tools
    but also the relative strengths of different tools compared to each other. 
    
    Along with this paper, we provide an online repository\footnote{\url{https://github.com/MarbleHE/SoK}} 
    that includes Docker images for all the tools we evaluate, our automated benchmarking framework, and the example applications.
    Additionally, it includes further benchmarks and tool descriptions that we could not include in this paper due to space limitations.

    We conclude our paper with a discussion of the current state of FHE and FHE tools.
    We discuss applications for which FHE is likely practical today and show gaps between 
    state-of-the-art results and what non-expert users can realistically implement.
    Based on the insights gained through our study, we highlight successes in the FHE tool space 
    and  identify gaps that remain to be addressed.
    Finally, we discuss a possible road map for the next generation of FHE tools.

\subsection{Related Work}

This work is, to the best of our knowledge, the first to survey and study the space of FHE compilers.
While previous work has considered FHE libraries~\cite{Acar2018}, they did so by contrasting different FHE schemes and their implementations and considered only a small subset of the tools we consider.
This paper is similar in nature to an SoK by Hastings et al.~\cite{Hastings2019-qw}, 
    which analyzed tools in the MPC domain.
    Practical MPC constructions have been an active research area since the 1980s. 
    As a consequence, these tools are more mature and more integrated into the research process.
 In their work, Hastings et al. focus on the usability aspects of MPC tools and did not consider performance.
    In contrast, we focus on analyzing the expressiveness and performance of the existing FHE tools
     and provide recommendations to  developers in choosing the correct tool for their target application.

\pagebreak
\section{Fully Homomorphic Encryption}
A \emph{homomorphic} encryption scheme is an encryption scheme where there exists a homomorphism between operations on the plaintext and operations on the ciphertext. 
For example, the Paillier encryption scheme~\cite{Paillier1999a} is \emph{additively} homomorphic: the product of two ciphertexts decrypts to the sum of their plaintexts.
Meanwhile, textbook RSA~\cite{Rivest1978} is \emph{multiplicatively} homomorphic: the product of two ciphertexts decrypts to the product of their plaintexts.
While such \emph{partially} homomorphic schemes have existed since the 1970s, \emph{fully} homomorphic encryption schemes, i.e., schemes that are homomorphic regarding both addition and multiplication, were an open problem until recently.
While first proposed shortly after the introduction of public-key cryptography in the 1970s~\cite{Rivest1978a},
proposed solutions could support only limited combinations of operations (e.g., additions plus a single multiplication~\cite{Boneh2005}).
    Here, we provide a brief overview over the development of modern FHE schemes, deferring more in-depth descriptions of selected FHE schemes to the appendix~(§\ref{fhe-appendix}).
    \fakeparagraph{Foundations of practical FHE.}
Modern FHE schemes date back to 2009 when Craig Gentry presented the first feasible \ac{fhe} construction~\cite{Gentry2009-zi}.
  While the original scheme had impractically large constant overheads,
follow-up work improved upon the scheme, enabling a first implementation~\cite{Gentry2011-kk}.

All modern schemes follow the general approach laid out by Gentry's first scheme: In these schemes,
    public keys are values that cancel out to zero when combined with the secret key $\sk$.
    Encryption multiplies the public key $\pk$ with a random number $a$ and adds the message $m$.
    For two ciphertexts $x_i = \pk *  a_i + m_i$, addition and multiplication trivially translate to the equivalent operations on the plaintexts, i.e., 
    $x_0 * x_1 = m_0 * m_1 +  \pk *  (a_0 *  x_1 + a_1 * m_0)$ which decrypts to $m_0 * m_1$ when combined with the secret key.
    However, for a secure system, \emph{noise} must be added to public keys and ciphertexts.
    As long as the noise  $e$ is sufficiently small, $m + e'$ can be rounded to the correct value and applying the secret key recovers $m$.
    During homomorphic operations, the noise in the ciphertext grows. 
    While this effect is negligible during additions, 
    multiplying two ciphertexts introduces significantly more noise.
    This limits computations to a (parameter-dependent) number of consecutive multiplications (multiplicative \emph{depth}) before decryption fails.
    This limitation can be circumvented using \emph{bootstrapping}, which resets the noise level
    of a ciphertext to a fixed lower level by homomorphically evaluating the decryption circuit with an
    encrypted secret key as input.
    However, the decryption circuit needs to be sufficiently low-depth to allow at least one additional multiplication before needing to bootstrap again.
    
    \fakeparagraph{Second Generation Schemes.}
While the first generation of FHE presented a significant academic breakthrough, it was too inefficient
(e.g., around \SI{30}{\minute} needed to compute a single homomorphic multiplication) 
to truly enable practical applications of FHE.
    In response, a second generation of schemes like the \acf{bgv}~\cite{Brakerski2014-uq} and \acf{bfv}~\cite{fan2012somewhat, brakerski2012fully} schemes evolved.
    In order to overcome the performance penalties of bootstrapping, they introduced the concept of \emph{leveled} homomorphic encryption.
    Here, the parameters are chosen sufficiently large to evaluate the entire computation without bootstrapping.
    While there is a cut-off point after which bootstrapping is more efficient, this is unlikely to be reached by most programs. 
    In addition, they introduced support for \acf{simd}-style \emph{batching}.
    This exploits the fact that the plaintext space is a ring of polynomials with many coefficients.
    Using the Chinese Remainder Theorem~\cite{iliashenko2019optimisations}, this can be reinterpreted as many different
    independent slots and many different messages (usually $2^{13}$--$2^{16}$) can be packed into a single ciphertext.
    Automorphisms additionally enable 
    homomorphically executable rotations between slots~\cite{halevi2018faster}.

    The \acf{ckks} scheme~\cite{cheon2017homomorphic} introduces a further optimization, considering 
     homomorphic encryption for \emph{approximate} numbers. 
     While it follows a very similar construction to \ac{bgv}, it is formally speaking not an FHE scheme since the result is only approximately the same as the equivalent plaintext operation,
     which can introduce subtle issues in practice.
     However, this relaxation has led to an extremely efficient scheme.
    \ac{ckks} is designed primarily for computations with fixed-point numbers, i.e., a number $x$ is represented as $m = \nearest{x*\varDelta}$ for \emph{scale} $\varDelta$, usually a large integer.
    \ac{ckks} introduces a homomorphic rounding operation to reduce the scale homomorphically, avoiding overflow issues.

\setlength{\skip\footins}{2mm}
    \fakeparagraph{Third Generation Schemes.}
More recently, a third generation of FHE schemes, based on the \acf{gsw} scheme~\cite{gentry2013homomorphic}, has emerged.
    These schemes mostly abandon batching and leveled HE and instead focus on fast bootstrapping.
For example, implementations of the \acl{cggi} \acused{cggi} (CGGI\footnote{The \ac{cggi} scheme is more commonly known
as TFHE, however we refer to it by the author initials in order to avoid confusion with the TFHE library.})
scheme~\cite{Chillotti2016Dec,Chillotti2017-yd} can perform bootstrapping in less than \num{0.1}~seconds,
while bootstrapping for \ac{bfv} or \ac{bgv} usually takes several minutes even in efficient implementations.
    While initially limited to binary settings, recent follow-up work~\cite{Chillotti2020-ia} extends this to arithmetic circuits.
    However,
    fast bootstrapping is incompatible with batching, introducing a trade-off between latency and throughput when compared to second-generation schemes.

\fakeparagraph{FHE and MPC.}
Finally, we briefly consider FHE in the wider context of secure \ac{mpc}.
    While FHE could be used to realize many 2-party \ac{mpc} protocols, it does not by default offer \emph{circuit privacy}, i.e., does not hide the function being computed.
    Where desired, this is usually addressed in practice via \emph{noise flooding}~\cite{Gentry2009-zi}, i.e., adding large noise to the final result before returning it to the client.    
    FHE can also be extended to multi-party or multi-key settings.
    In multi-party FHE, different entities generate a public key and shares of a secret key~\cite{Mouchet2020-pz}.
    In multi-key FHE, each entity independently generates their secret and public key~\cite{Chen2019-ja}.
    There are also hybrid schemes that combine FHE and \ac{mpc}~\cite{juvekar2018} or different FHE schemes~\cite{Boura2018-wj}. 
    We only consider the two-party FHE-only client-server setting, but many of the concepts transfer directly to the other settings.

\section{What makes developing FHE applications challenging?}

The intricacy of the underlying schemes still limits developing FHE-based applications predominantly to experts. Each scheme presents a new set of configurations and performance tradeoffs, and achieving state-of-the-art results requires a high familiarity with the underlying schemes. 
In addition, FHE imposes a fundamentally different programming paradigm, 
not only because of the need for data-independent programs but also because efficient solutions frequently require complex vectorization approaches.

Throughout the last decade, a significant amount of \emph{folklore} knowledge around optimization methods and best practices has been built up in the FHE community.
However, these techniques and insights are often scattered across the literature or only referred to in passing.
As a result, there is a vast gap between state-of-the-art performance results and what non-experts can achieve themselves.

In this section, we provide an overview of the key engineering challenges that developers face today.
The community is starting to identify these accessibility issues as a major roadblock to the broader adoption of FHE.
Recent works are trying to address these challenges by proposing higher-level interfaces, better abstractions, and automated 
optimizations.
There will most likely always be specific applications that impose additional challenges requiring expert input.
However, improved tools can benefit a variety of common application patterns and help ease the path to FHE for many applications.

\subsection{Parameter Selection}
Selecting secure and efficient instantiations of the underlying cryptographic problems is hard for most encryption schemes.
In standard public-key cryptography, we circumvent this by standardizing particular instantiations, e.g., selecting certain elliptic curves,
to avoid security issues arising when the underlying hardness assumptions do not hold for poor choices.

FHE introduces the additional challenge of computation-specific parameters.
More complex computations require larger plain- and ciphertext moduli to avoid overflow or noise issues.
However, as these parameters increase, the \ac{lwe} problem that security is based on for most schemes becomes easier, and the dimension of the problem space (i.e., polynomial degree) must be increased to compensate.
As a result, we cannot standardize a single set of secure parameter choices.
Instead, the standardization effort~\cite{HomomorphicEncryptionSecurityStandard} aims to provide a conservative estimate of the security of different combinations of moduli and dimensions. 
However, since this does not address efficiency, parameter selection remains an issue in developing FHE-based applications.

The time to evaluate homomorphic operations, for a given polynomial degree $n$, is roughly proportional to the ciphertext space modulus $q$, and a smaller $q$ also gives higher security.
Therefore, we want to select the smallest $q$ that still correctly decrypts the computation result.
However, effectively computing this minimal $q$ remains an open challenge.
While formal analyses of the ciphertext noise growth exist for a variety of schemes, these worse-case analyses are frequently too conservative, giving parameters many times larger than the experimentally determined optimum~\cite{Costache2019-jd}.
Also, the plaintext space modulus $t$ required to avoid overflows depends on the size of the actual inputs, which likely come from a smaller subset of $\Z_t$ in practice.
Here, again, worst-case analyses lead to impractically conservative parameters.
Instead, the community's accepted method is to incrementally decrease $q$ until the computation (on some representative input values) fails to decrypt correctly, then choosing the previous $q$ plus some ``safety margin" determined by experience.

\vspace{5pt}
\subsection{Encoding}

    With encryption schemes like AES or protocols like TLS, developers do not generally have to consider the plaintext spaces of the underlying encryption schemes.
    As long as a message can be serialized into a binary string, only padding concerns arise.
    However, in FHE, the semantics of the plaintext space determine the effect of the homomorphic computations.
    These semantics, however, frequently do not match the intended application semantics exactly.
    While this is already a concern in traditional programming, with floating-point accuracy errors or integer overflows, 
    FHE introduces a significantly stronger deviation from the `ideal' computation model.

    For example, while we generally consider $\Z_t$ as the message space for most schemes, most support additional, more complex spaces. 
    For example, \ac{bgv} supports Galois Fields $\mathbf{GF}(2^d)$ which can be used to efficiently realize AES-FHE transciphering, i.e., converting a standard AES ciphertext to an FHE ciphertext given an encryption of the AES key~\cite{Gentry2012b}.

    Conceptually, binary plaintext spaces (i.e., $\Z_2$) are the easiest to work with since the semantics of homomorphic computations directly correspond to binary circuits. 
    However, working directly with binary circuits is complicated as even trivial functions like addition and
    multiplication of bit-wise encoded integers require complex algorithms (e.g., Sklansky or Kogge-Stone adders) to
    implement arithmetic operations efficiently.
    Therefore, the conceptual ease-of-use is negated by a significant engineering overhead for even simple algorithms.

    While using advanced encoding schemes will most likely remain predominantly an expert technique, 
    existing FHE tools have already shown that they can be employed automatically to some extent.
    For example, nGraph-HE~\cite{Boemer2019-mt} also uses the imaginary part of the CKKS message space 
    when no ciphertext multiplications are required, roughly doubling throughput.

\subsection{Data-Independent Computation}
    Virtually all standard programming paradigms rely on some form of data-dependent execution branching.
    Traditional iterative programming relies heavily on if/else statements and loops, 
    and even functional programming requires data-dependent branching to terminate recursion.
    
    FHE computations, on the other hand, are by definition data-independent, or they would violate the privacy guarantees.
    Therefore, FHE computations are frequently conceptualized as \emph{circuits}, i.e., gates (or operations) connected by wires, 
    where the execution follows the same steps, no matter what values the input has.
    While it is possible to emulate, e.g., if/else branches by calculating the result for both branches and performing a multiplexing selection afterward, this requires evaluating both branches.
    Simulating (bounded) dynamic-length loops could be achieved by following a similar approach; however, this quickly becomes infeasible in practice.
    
    In addition, many schemes offer the best performance when using integer plaintext spaces ($t \gg 2$).
    These \emph{arithmetic circuits} are no longer Turing-complete and are instead limited to computing polynomial functions.
    However, many applications, including neural-network inference, can be approximated very well.
    Therefore, a significant part of developing an FHE-based solution is to consider first whether there exists a polynomial approximation for the task to be performed.
    Sometimes, this even requires completely switching the approach, e.g., standard algorithms for genomic sequence analysis are not suitable for polynomial approximation, but alternative approaches exist that can be expressed much more easily~\cite{Lauter2014-tv}.

\vspace{5pt}
\subsection{\ac{simd} Batching} \label{subsec:simd-batching}
    One of the major breakthroughs in achieving practical performance in FHE-based solution was the introduction of \emph{batching} or \emph{packing} in second-generation schemes,
  i.e., allowing one to pack many different messages into a single ciphertext.
     The resulting \ac{simd} parallelism can trivially be used to improve throughput by packing many different inputs into a single computation run.

     However, many FHE applications are limited in their practicality by latency, i.e., non-amortized runtime.
     State-of-the-art FHE-based solutions virtually always apply batching inside a computation, even on a single instance of the input.
      Exploiting SIMD batching to reduce latency requires novel programming paradigms and algorithms that do not have equivalents outside FHE.
    For example, matrix-vector-products can be expressed more efficiently if we encode each of the matrix diagonals into a SIMD vector~\cite{Halevi2014-cb}, rather than row- or column-wise. %

    SIMD batching is, for those schemes that support it, potentially the most important optimization technique, as the large size of the vectors can lead to runtime improvements of many orders of magnitude.
    However, it is also one of the more complex techniques, requiring a deep understanding of both the application and the performance-tradeoffs of the FHE scheme in question. 
While some domains, such as machine learning, are inherently heavily vectorized and can therefore be automatically transformed into SIMD-friendly forms, this remains an open problem for more general applications.

\subsection{Ciphertext Maintenance}
    Different schemes use a variety of solutions to manage the growth of the ciphertext noise during homomorphic computations.
    However, virtually all schemes feature some form of \emph{ciphertext maintenance} operations.
    These are operations like relinearization, mod-switching/rescaling or bootstrapping that must be called explicitly by the developer in order to manage the noise growth optimally.
    For example, while one might be tempted to apply relinearization immediately after each multiplication, doing so is suboptimal.
    This is most obvious for the last multiplication in a computation: with no further multiplications following, the benefit of reducing future noise growth is lost.
    Similar issues appear when considering when to rescale in the CKKS scheme.
    
    Bootstrapping is frequently not efficient when a leveled approach can be used.
    However, there are some applications for which it is the more suitable approach.
    In general, there is a continuum of choices between the minimal parameters that allow only a single operation before bootstrapping is needed and the (potentially infeasibly large) parameters required to execute the entire computation without bootstrapping.
    
    One of the major advantages of the CGGI scheme is that it inherently relies on bootstrapping to realize each
    operation. Therefore, it removes the developer's burden to consider parameters and bootstrapping.
    However, it is worth noting that a leveled version of the scheme is, in fact, faster for certain applications, once again demonstrating a trade-off between simplicity and performance.
   
    While a variety of tools have included automatic ciphertext maintenance~\cite{crockett2018alchemy,Viand2018-cs,Carpov2015-ok}, these were usually naive heuristics that did not improve performance.
     Developing efficient strategies is difficult because there are usually multiple degrees of freedom.
     For example, for the rescaling operations in CKKS one needs to consider both what scale to rescale to and where to insert the operations.
     Recently, however, there have been increasing efforts to automate this process~\cite{Dathathri2019-vu}.

\section{Survey Methodology}
\label{overview}

    We split our analysis of the FHE tool space into two parts.
    First, we present an extensive survey of existing tools and highlight their features and characteristics.
    Second, we consider these tools in practice by experimentally evaluating them across a range of case study applications, 
    contrasting usability, expressiveness, and performance.
    We combine our quantitative performance analysis with a qualitative assessment, describing the challenges of developing applications in the different tools.

    The secure computation ecosystem includes many different types of tools.
    On the low-level side, there are math libraries that simplify building implementations of FHE schemes, e.g., by efficiently implementing techniques useful for general lattice cryptography. %
    Then, there are FHE libraries that implement specific schemes and offer slightly higher-level APIs, e.g., \texttt{keygen}, \texttt{encode}, \texttt{encrypt}, \texttt{add}, \texttt{mult}.
    Finally, there are compilers that abstract aspects like parameter selection, encryption and decryption by offering a higher-level language that developers can use to specify their computation.

    In our survey, we consider FHE libraries and compilers.
    While some of the underlying math libraries provide implementations of FHE schemes as examples~\cite{crockett2016lambdaolambda}, we consider only tools that natively offer an API for FHE operations.
    Throughout the last decade, there has been significant development in schemes and implementations, with some being discarded or replaced for security or efficiency reasons.
    We only consider tools based on schemes that are currently still considered viable candidates (i.e., \ac{bfv}, \ac{bgv}, \ac{ckks}, or \ac{gsw}-based constructions) 
    and consider only the latest version of each tool, including ``spiritual successors" where they exist. 
    We also consider only unique implementations, i.e.,  we do not list wrappers or ports of existing tools.
    FHE techniques are used internally in several MPC protocols, and there are a variety of tools that specifically target hybrid protocols combining FHE and MPC~\cite{juvekar2018}.
    However, for this survey, we consider only tools that support using purely FHE, requiring no interaction during the computation itself.

We focus on three design aspects: 
\emph{(i)}~settings and configurations, e.g., which input languages or schemes a tool supports;
\emph{(ii)}~features and optimizations, e.g., support for batching or automated parameter selection;
\emph{(iii)}~accessibility, e.g., documentation and examples.

    In our experimental evaluation, we consider a selection of tools in more detail.
    Through using the tools to implement different case study applications, we can provide an in-depth analysis of their usability and expressiveness in practice.
    In addition, our benchmarks allow us to study not only the overall performance of FHE for these applications 
    but also the relative strengths of different tools compared to each other.
    We select three applications that represent different domains of FHE-based computation.
    Our first application is a risk score calculation that requires comparisons and, therefore, binary circuit emulation.
    Second, we consider a statistical $\chi^2$-test, in a formulation that simplifies it to polynomial functions over integers.
    Finally, we consider machine learning, specifically neural network inference, for a range of network architectures.
    We evaluate these applications across the different tools and report on usability, expressiveness, and performance.

\begin{table*}[]
    \centering
    \begin{tabular}{@{}llccccccccc@{}}
        \toprule
        \multirow{2}{*}[-0.2em]{Name}
        & \multirow{2}{*}[-0.2em]{\begin{tabular}[c]{@{}l@{}}Input Lang.\end{tabular}}
        & \multicolumn{3}{c}{{\scriptsize Supported Schemes}}
        & \multicolumn{2}{c}{{\scriptsize Features}}
        & \multicolumn{3}{c}{{\scriptsize Accessibility}}
        & \multicolumn{1}{l}{\multirow{2}{*}[-0.2em]{\begin{tabular}[c]{@{}l@{}}Last Major\\Update \end{tabular}}}                                                                                                                                                                         \\
        \\[-1.6em] %
        \cmidrule(lr){3-5} \cmidrule(lr){6-7} \cmidrule(lr){8-10}
        &                                                                        & \multicolumn{1}{l}{BFV} & CKKS    & GSW    & \multicolumn{1}{l}{Bootstrapp.} & Levels    & \multicolumn{1}{l}{Code} & Ex.         & Doc.       & \multicolumn{1}{l}{} \\ 
        \midrule
        concrete (\cite{Chillotti2020-ll}) & Rust	& \Circle & \Circle & \CIRCLE   & \CIRCLE & \Circle & \CIRCLE & \LEFTcircle & \CIRCLE & 11/2020 \\
        FHEW (\cite{Ducas2015-ei})             & \cpp                                                                   & \Circle                  & \Circle & \CIRCLE & \CIRCLE                         & \Circle & \CIRCLE                 & \Circle     & \Circle     & 05/2017              \\
        FV-NFLlib (\cite{CryptoExperts2016-yq}) & \cpp								& \CIRCLE			&\Circle & \Circle  & \Circle & \CIRCLE & \CIRCLE & \LEFTcircle & \Circle & 07/2016 \\
        HEAAN (\cite{cheon2017homomorphic})     & \cpp                                                                   & \Circle                  & \CIRCLE & \Circle & \CIRCLE                         & \CIRCLE & \CIRCLE                 & \LEFTcircle & \Circle     & 09/2018              \\
        HElib (\cite{Halevi2014-bh})            & \cpp                                                                   & \CIRCLE                  & \CIRCLE & \Circle & \CIRCLE                         & \CIRCLE & \CIRCLE                 & \CIRCLE     & \CIRCLE     & 12/2020             \\
        lattigo (\cite{Mouchet2020-kh}) & Go 	& \CIRCLE			&\CIRCLE & \Circle   & \Circle & \CIRCLE & \CIRCLE & \CIRCLE & \Circle & 12/2020\\
        \Palisade (\cite{polyakov2019palisade}) & \cpp                                                                   & \CIRCLE                  & \CIRCLE & \CIRCLE & \CIRCLE                         & \CIRCLE & \CIRCLE                 & \CIRCLE     & \CIRCLE     & 04/2020              \\
        SEAL (\cite{sealcrypto})                & \cpp, .NET                                                             & \CIRCLE                  & \CIRCLE & \Circle & \Circle                         & \CIRCLE & \CIRCLE                 & \CIRCLE     & \CIRCLE     & 08/2020              \\
        TFHE (\cite{TFHE})                      & \cpp                                                                   & \Circle                  & \Circle & \CIRCLE & \CIRCLE                         & \Circle & \CIRCLE                 & \CIRCLE     & \CIRCLE     & 05/2017              \\    
        \midrule
        cuFHE (\cite{Vernam_Group2018-mf})   & \cpp, Python 	& \Circle			&\Circle & \CIRCLE   & \CIRCLE & \Circle & \CIRCLE & \LEFTcircle & \Circle & 08/2018 \\
        nuFHE (\cite{NuCypher2019-dc})  & \cpp, Python 	& \Circle			&\Circle & \CIRCLE   & \CIRCLE & \Circle & \CIRCLE & \CIRCLE & \CIRCLE & 07/2019\\
        \bottomrule
    \end{tabular}
    \caption{
        Overview of existing \acs{fhe} CPU-targeting (top) and GPU-targeting (bottom) libraries. 
         Note that similar schemes are summarized into categories, e.g., BFV/BGV as BFV and CGGI/TFHE/FHEW as GSW. 
    }
    \label{tbl:overview-libraries}
\end{table*}

\section{\acs{fhe} Libraries} \label{subsec:lower_level_libs}

    \ac{fhe} libraries implement the underlying cryptographic operations of an \ac{fhe} scheme and expose a higher-level API.
    They minimally provide key generation, encryption, decryption, homomorphic addition, and multiplication interfaces.
    In practice, however, library APIs often include dozens of additional functionalities for ciphertext maintenance and manipulation.

    Using these libraries generally requires a deep understanding of the underlying scheme and its supported operations.
    While many libraries include powerful advanced features that can significantly improve performance, %
    developers must employ them manually while ensuring correctness and efficiency.
        
    In \Cref{tbl:overview-libraries}, we present an overview of \ac{fhe} libraries and list supported languages, schemes, features, and accessibility aspects.
    We group schemes into families of related schemes for conciseness and consider support for bootstrapping and leveled-\acs{fhe}.
    For accessibility, we consider whether an implementation (Code) is available, whether examples (Ex.) describe usage (\CIRCLE) or usage can be inferred from, e.g., tests (\LEFTcircle), and whether or not API documentation (Doc.) is available.
    Finally, we give a rough indication of age and activity by giving the date of the last release or major update.

    Due to space constraints, we present only a small subset in more detail.
    We start by discussing HElib, SEAL, and Palisade, which appear to be the most active and widely supported libraries.
    We also discuss TFHE here since it is used by some of the compilers we evaluate.
    Finally, we discuss performance differences and briefly discuss the remaining libraries.

\subsection{\acs{helib}}
The \acf{helib}, presented in 2013 by Halevi and Shoup, was the first FHE library~\cite{Halevi2014-cb}. 
The library is implemented in \cpp and uses the NTL library~\cite{Shoup2016} for the underlying mathematical operations.
While it initially only implemented the \ac{bgv} scheme, more recent releases of this library also support the \ac{ckks} scheme.
The library offers leveled FHE operations and, for \ac{bfv}, also supports bootstrapping~\cite{Halevi2015}.
The source code is available under the Apache License~v2.0, and includes extensive examples.
In addition to the standard documentation, several reports describing the design and algorithms of \ac{helib}~\cite{Halevi2014-bh,Halevi2015, Halevi2020-ok} are available.

    \subsection{PALISADE}
    \Palisade, first released in 2014, is developed primarily by NJIT and Duality Technologies~\cite{polyakov2019palisade}.
    It is implemented in \cpp and optionally uses the NTL library~\cite{Shoup2016} to accelerate underlying mathematical operations.
    \Palisade supports a wide range of schemes, including \ac{bfv}, \ac{bgv}, \ac{ckks}, and \ac{cggi}.
    In addition, it supports multi-party extensions of certain schemes and other cryptographic primitives like proxy re-encryption and digital signatures.
    The library offers both leveled and bootstrapped operations, where supported by the scheme.
    \Palisade's source code is available under a BSD 2-clause license and includes examples and documentation.

\subsection{\acs{seal}}\label{subsec:rel_work_seal}

The \acf{seal}, first released in 2015, is developed by Microsoft Research~\cite{sealcrypto}.
It is implemented in {\cpp}, with an official wrapper for \texttt{.NET} languages (e.g., C\texttt{\#}). %
\ac{seal} is thread-safe and heavily multi-threaded itself.
It implements the \ac{bfv} and \ac{ckks} schemes, with a majority of the API being common to both.
\acs{seal} offers leveled FHE operations and does not implement bootstrapping for either scheme.
Earlier versions of \ac{seal} included automated parameter selection %
based on estimating the noise growth~\cite{player2018parameter}.
Since the estimated parameters were frequently non-competitive, this feature was removed.
However, \ac{seal} still ensures that the chosen parameters offer 128-bit security.
The source code is available under an MIT license, is well documented, and includes a wide range of examples for both schemes. 
In addition, there are several demo applications (e.g., AsureRun~\cite{Microsoft2019-qj}) that demonstrate more complex use cases.

\subsection{\acs{tfhe}}

The \acf{tfhe} was proposed in 2016 by Chillotti et al.~\cite{Chillotti2016Dec} and can be considered the successor of the \fhew library~\cite{Ducas2015-ei}.
It is implemented in \cpp and supports a variety of different libraries for underlying FFT operations.
\ac{tfhe} is based on the \ac{cggi} scheme and offers gate-by-gate bootstrapping with significantly reduced bootstrapping times,
resulting in times of less than \SI{0.1}{\sec} compared to \SI{6}{\min} for bootstrapping in the \acs{helib} library.
\ac{tfhe} implements a variety of logic gates like \OrGate, \NorGate, \MuxGate that are generally implemented more efficiently than naive constructions from \Xor and \AndGate would be.
However, the library provides no assistance with building more complex logic circuits like efficient comparators and adders.
    \ac{tfhe}'s source code is available under the Apache License~v2.0 and includes examples and documentation.

\subsection{Other Libraries}
In addition to the libraries we discussed above, we considered a large variety of other libraries~\cite{cheon2017homomorphic, Ducas2015-ei, CryptoExperts2016-yq, Mouchet2020-kh, Chillotti2020-ll}.
We also conducted a series of microbenchmarks to compare how different implementations of the same scheme perform.
However, due to space considerations we refer to our accompanying online repository for details.
Finally, GPU-based libraries like cuFHE~\cite{Vernam_Group2018-mf} and nuFHE~\cite{NuCypher2019-dc} can offer significant speedups, improving the already fast \ac{tfhe} bootstrapping times by around two orders of magnitude.
However, as GPUs remain considerably more expensive and less common in enterprise datacenters, these speedups must be considered in context.

\begin{table*}[t]
    \centering
    \begin{tabular}{@{}llccclccccccccc@{}}
        \toprule
        \multirow{2}{*}[-0.25em]{Name}
        & \multirow{2}{*}[-0.25em]{\begin{tabular}[c]{@{}l@{}}Input \\ Lang.\end{tabular}}
        & \multicolumn{3}{c}{{\scriptsize Schemes}}
        & \multirow{2}{*}[-0.25em]{\begin{tabular}[c]{@{}l@{}}Ptxt. \\ Space\end{tabular}}
        & \multicolumn{4}{c}{{\scriptsize Features \& Optimizations}}
        & \multicolumn{3}{c}{{\scriptsize Accessbility}}
        & \multicolumn{1}{l}{\multirow{2}{*}[-0.25em]{\begin{tabular}[c]{@{}l@{}}Last Major\\ Update\end{tabular}}}                                                                                                                                                                                                                                                                                                                                                                     \\
        \\[-1.6em] %
        \cmidrule(lr){3-5} \cmidrule(lr){7-10} \cmidrule(lr){11-13}
        &                                                                         & \multicolumn{1}{l}{BFV} & \multicolumn{1}{l}{CKKS} & \multicolumn{1}{l}{GSW} &               & \multicolumn{1}{l}{SIMD} & \multicolumn{1}{l}{Params.} & \multicolumn{1}{l}{Ctxt. Mnt.} & \multicolumn{1}{l}{$\times$ Depth} & \multicolumn{1}{l}{Code} & \multicolumn{1}{l}{Ex.} & \multicolumn{1}{l}{Doc.} & \multicolumn{1}{l}{} \\
        \midrule
        \acs{alchemy} (\cite{crockett2018alchemy})  & Haskell                                                                 & \CIRCLE                  & \Circle                  & \Circle                  & Arithm.       & \LEFTcircle              & \LEFTcircle                & \LEFTcircle                                        & \Circle                            & \CIRCLE                 & \LEFTcircle             & \Circle                   & 02/2020            \\
        Cingulata (\cite{Carpov2015-ok})            & \cpp                                                                    & \CIRCLE                  & \Circle                  & \LEFTcircle                  & Binary        & \Circle                  & \CIRCLE                    & \LEFTcircle                              & \CIRCLE                            & \CIRCLE             & \CIRCLE                 & \Circle                   & 11/2019              \\
        E\textsuperscript{3} (\cite{chielle2018e3}) & \cpp                                                                    & \CIRCLE                  & \Circle                  & \CIRCLE                  & Both 		   & \LEFTcircle                  & \Circle                    & \LEFTcircle                                       & \Circle                            & \CIRCLE                 & \CIRCLE                 & \CIRCLE               & 09/2020              \\
        EVA (\cite{Dathathri2019-vu})               & Python                                                                  & \Circle                  & \CIRCLE                  & \Circle                  & Arithm.       & \CIRCLE                  & \CIRCLE                    & \CIRCLE                                            & \Circle                            & \CIRCLE                 & \CIRCLE                      & \Circle                        & 11/2020                   \\
        Marble (\cite{Viand2018-cs})                & \cpp                                                                    & \CIRCLE                  & \Circle                  & \Circle                  & Both       & \LEFTcircle              & \LEFTcircle                    & \LEFTcircle                            & \Circle                            & \CIRCLE                 & \CIRCLE                 & \Circle                   & 10/2018              \\
        \Ramparts (\cite{Archer2019-iy})            & Julia                                                                   & \CIRCLE                  & \Circle                  & \Circle                  & Arithm.       & \Circle                  & \CIRCLE                    & \LEFTcircle                                       & \CIRCLE                            & \Circle                 & --                      & --                        & --                   \\
        \midrule[0.01em]
        CHET (\cite{dathathri2018chet})             & \cpp                                                                    & \Circle                  & \CIRCLE                  & \Circle                  & Arithm.       & \LEFTcircle              & \CIRCLE                    & \LEFTcircle                                      & \Circle                            & \Circle                 & --                      & --                        & --                   \\
        nGraph-HE (\cite{Boemer2019-mt})            & Python                                                                  & \CIRCLE                  & \CIRCLE                  & \Circle                  & Arithm.       & \CIRCLE                  & \Circle                    & \LEFTcircle                                         & \CIRCLE                            & \CIRCLE                 & \CIRCLE                 & \CIRCLE                   & 08/2019              \\
        SEALion (\cite{van2019sealion})             & Python                                                                  & \CIRCLE                  & \Circle                  & \Circle                  & Arithm.       & \LEFTcircle              & \CIRCLE                    & \LEFTcircle                                         & \Circle                            & \Circle                 & \CIRCLE                      & \Circle                        & 01/2019                   \\
        \bottomrule
    \end{tabular}
    \caption{Overview of existing general-purpose \acs{fhe} compilers (top) and those specializing on machine learning (bottom). 
        Note that similar schemes are summarized into families, e.g., BFV/BGV as BFV and CGGI/TFHE/FHEW as GSW.
    }
    \label{tbl:overview-compilers}
\end{table*}

\begin{table}[]
    \centering
    \begin{tabular}{llllllll}
        & \rot{FHEW} & \rot{HEAAN} & \rot{HElib} & \rot{\Palisade} & \rot{SEAL} & \rot{TFHE} \\
        \toprule
        \acs{alchemy}        & \Circle    & \Circle     & \Circle     & \Circle         & \Circle    & \Circle    \\
        Cingulata            & \Circle    & \Circle     & \Circle     & \Circle         & \Circle    & \CIRCLE    \\
        E\textsuperscript{3} & \CIRCLE    & \Circle     & \CIRCLE     & \Circle         & \CIRCLE    & \CIRCLE    \\
        EVA                  & \Circle    & \Circle     & \Circle     & \Circle         & \CIRCLE    & \Circle    \\
        Marble               & \Circle    & \Circle     & \CIRCLE     & \Circle         & \CIRCLE    & \Circle    \\
        \Ramparts            & \Circle    & \Circle     & \Circle     & \CIRCLE         & \Circle    & \Circle    \\
        \midrule[0.01em]
        CHET                 & \Circle    & \CIRCLE     & \Circle     & \Circle         & \CIRCLE    & \Circle    \\
        nGraph-HE     & \Circle    & \Circle     & \Circle    &\Circle       & \CIRCLE & \Circle \\
        SEALion              & \Circle    & \Circle     & \Circle     & \Circle         & \CIRCLE    & \Circle    \\
        \bottomrule
    \end{tabular}
    \caption{Use of existing \acs{fhe} libraries by \ac{fhe} compilers. Note that \acs{alchemy} implements BGV internally using the \Lol lattice cryptography library, and Cingulata also includes a custom implementation of BFV.}
    \label{tbl:overview-dependencies}
\end{table}

\section{\acs{fhe} Compilers} \label{subsec:fhe_compiler}

    This section provides an overview of existing \ac{fhe} compilers, 
    i.e., tools that provide a high-level abstraction to develop \ac{fhe}-based applications, so developers do not have to deal directly with homomorphic operations on ciphertexts.
    These tools generally manage key setup, encryption, decryption, and ciphertexts maintenance operations in the background.
    The term \emph{compiler} is used loosely in the context of \ac{fhe}, as some function more like interpreters or libraries to link against.
    
    In \Cref{tbl:overview-compilers}, we provide an overview of the various \ac{fhe} compilers and their properties.
    \ac{fhe} compilers can roughly be divided into generic tools for general purpose use
    and tools that target specific applications.
    In the latter category, we see compilers targeted at building \ac{ml} applications.
    In addition to supported schemes, which we again group for conciseness, 
    we also consider the plaintext spaces supported by the tool.  
    Even when the underlying scheme and implementation support different plaintext spaces, 
    compilers generally only target binary or arithmetic plaintext spaces.
   
    We consider a wide range of features and generally differentiate between three states indicating full support (\CIRCLE), partial support (\LEFTcircle), or no support (\Circle).
    For SIMD-Batching (SIMD), we differentiate between tools that merely enable batching and those that actively assist in working with vectorized data.
    Similarly, we differentiate between manual, partially assisted, and fully automated parameter selection (Params.).
    While all tools include some form of automated ciphertext maintenance operations (Ctxt. Mnt.), 
    we segment tools into those that use naive heuristics and those using more advanced strategies.
    Additionally, we note whether or not tools try to reduce the multiplicative depth ($\times$ Depth) of the circuits they generate.
    For accessibility, we consider the same metrics as for libraries, i.e., whether an implementation (Code) is available, whether examples (Ex.) describe usage, and whether or not API documentation (Doc.) is available.
    Similarly, we again give a rough indication of age and activity by giving the date of the last release or major update.
    Where no source code is available to us, we have to omit these metrics (``--").    
    
    Finally, \Cref{tbl:overview-dependencies} associates compilers with the libraries they use.
    Here we can see \ac{seal} being targeted by a significantly larger number of compilers than any other library.   
    
    In the following, we introduce each compiler in more detail.

\subsection{\acs*{alchemy}}
\acf{alchemy} was proposed by Crockett et al. in 2017~\cite{Crockett2017}.
Input programs are specified in a special \ac{dsl} implemented in Haskell
and executed as arithmetic circuits using a custom \ac{bgv} implementation using the \Lol lattice crypto library~\cite{crockett2016lambdaolambda}.
While it supports SIMD batching, it does not offer an encoding/decoding API, making it difficult to use.
\ac{alchemy} automatically selects suitable parameters by statically tracking the upper bound of the ciphertext error but requires user-supplied modulus candidates.
However, this approach, based on type-level arithmetic, leads to excessively long compilation times and makes \ac{alchemy} impractical for complex programs.
While open-source, the minimal examples are insufficient to allow non-Haskell-experts to use the library, and it is therefore excluded from our experimental evaluation.

\subsection{Cingulata}
Cingulata (previously \emph{Armadillo}) was proposed in 2015 by Carpov et al.~\cite{Carpov2015-ok}.
The compiler takes {\cpp} code as input and generates a corresponding Boolean circuit. %
Cingulata implements the \ac{bfv} scheme directly, using
the Flint and Sage libraries for operations on polynomials.
We refer to this built-in BFV implementation as CinguBFV. %
Cingulata also supports the CGGI scheme via the TFHE library, but advanced optimizations are not supported in this mode.
Recent versions include CinguParam~\cite{Herbert2019-ci}, which automatically determines parameters for BFV.
Cingulata inserts relinearization operations naively but tries to reduce the circuit's multiplicative depth using
the circuit optimization tool ABC~\cite{mishchenko2018abc},
which was originally designed for hardware synthesis. %
However, follow-up work has introduced novel FHE-specific depth-reduction heuristics~\cite{carpov2017multi,Aubry2020-jy, Lee2020-uq}.
Cingulata's source code is available under the CeCILL license and includes many examples.

\subsection{\acl*{e3}}
The \acf{e3} framework was proposed by Chielle et al.~\cite{chielle2018e3} in 2018. %
\ac{e3} uses \cpp as its input language and
supports both arithmetic and boolean circuits in \ac{bfv}, \ac{bgv}, and \ac{cggi}. %
\ac{e3} supports SIMD operations but does not expose rotation operations, severely limiting the expressiveness.
Users must provide parameters as part of the configuration, and ciphertext maintenance operations are inserted naively.
It uses the Synopsys Design Compiler, a proprietary tool for hardware design, to try to reduce the circuit's multiplicative depth.
Internally, it supports a variety of libraries, including \acs{tfhe}, \fhew, %
\acs{helib}, and \ac{seal}. %
\ac{e3}'s source code is available online and includes both examples and documentation.

\subsection{\acs*{eva} \& \acs*{chet}} \label{subsec:eva_compiler}
The \acf{eva} was presented by Dathathri et al.~\cite{Dathathri2019-vu} in 2019. %
It introduces a novel input language explicitly designed for vector arithmetic 
and targets arithmetic circuits in \ac{ckks} using the SEAL library.
It is inherently batched and focuses on automating parameter selection and ciphertext maintenance.
The program is converted into a term graph, and during multiple passes, graph rewriting rules transform it, e.g., by inserting relinearization and rescaling operations at the optimal locations.
However, \ac{eva} does not consider depth-reducing transformations.
While EVA can be used for any (vectorized) application, the focus is primarily on neural network inference.
Towards this end, EVA includes and subsumes prior work in the form of the \ac{chet}~\cite{dathathri2018chet}, which focuses on optimizing matrix-vector operations.
\ac{eva} and its examples are available under the MIT license. \ac{chet}, however, is not.

\subsection{Marble}

Marble, presented by Viand et al. in 2018~\cite{Viand2018-cs} offers a high-level interface for FHE in \cpp by overloading built-in operators.
For arithmetic circuits, it targets BFV via the SEAL library, %
and for binary circuits, BGV as implemented in the HElib library. %
Marble exposes a batched version of the API, allowing relatively efficient implementation, but it requires that the developer provides a suitably vectorized program.
However, Marble provides only rudimentary parameter selection, inserts ciphertext maintenance operations naively, and does not apply any program optimizations.
While a version of Marble is available online, the available code supports only binary circuits.
Since Marble targets an outdated version of HElib and focuses on usability over optimizations, we do not include it in our experimental evaluation.

\subsection{Ramparts}
Ramparts was proposed in 2019 by Archer et al.~\cite{Archer2019-iy}.
It uses Julia, a language for interactive scientific computing, as its input language, 
and targets arithmetic circuits in \ac{bfv} using the \Palisade library.
Ramparts does not support batching, but includes noise-growth-estimation based parameter selection.
Ciphertext maintenance operations are inserted naively, but
 a symbolic simulator simplifies the circuit by applying sub-expression elimination, constant folding, and partial evaluation (e.g., loop unrolling, function inlining).
Ramparts is not publicly available. Therefore,  we were unable to include it in our experimental evaluation. %
However, Rampart's evaluation compares it against Cingulata and a baseline using \Palisade directly.
The evaluation showed significant performance benefits compared to Cingulata;
however, in exchange, Ramparts is limited to programs that can be expressed as polynomial functions and the symbolic evaluation approach significantly increases compilation times.

\subsection{nGraph-HE}
The nGraph-HE framework, proposed by Boemer et al.~\cite{boemer2019} in 2019, is based on Intel's nGraph \ac{ml} compiler~\cite{cyphers2018intel} 
and translates standard TensorFlow computations into arithmetic circuits in \ac{bfv} or \ac{ckks} using the SEAL library.
It enables inference on pre-trained models over encrypted inputs,
applying \acs{fhe}-specific optimizations (e.g., constant folding, SIMD-packing, and graph-level optimizations such as lazy rescaling and depth-aware encoding), and run-time optimizations (e.g., bypassing special plaintext values).
However, it inserts rescaling operations naively and requires the user to define the parameters.
In subsequent work~\cite{Boemer2019-mt}, nGraph-HE was extended to support non-polynomial activation functions.
However, these are computed in an interactive protocol with the client, which introduces significant latency and is out of scope for our study.
nGraph-HE is available under the Apache License~v2.0 and includes examples and documentation.

\subsection{SEALion}
The framework SEALion, proposed by Van Elsloo~\cite{van2019sealion} in 2019, 
exposes a custom Python API for specifying ML models, which are trained using TensorFlow.
    SEALion then enables inference over encrypted data using arithmetic circuits in \ac{bfv} using the SEAL library.
SEALion supports batching to increase inference throughput by performing inference over multiple data simultaneously but does not consider non-trivial batching optimizations.
Further, it features automatic parameter selection using a heuristic search algorithm to find an optimal parameter set.
However, it inserts ciphertext maintenance operations naively and does not consider depth-reducing optimizations.
SEALion is not currently publicly available; however, the authors shared their implementation with us, and the code includes well-commented examples.

\section{Experimental Evaluation}

    In the following, we present our experimental evaluation, where we investigate FHE compilers in more detail. %
    We use these tools to implement and benchmark selected case study applications.
    This allows us to provide an in-depth analysis of their usability and expressiveness in practice,
    and to compare the performance characteristics of current \ac{fhe} compilers.  
  
    Since there are no standardized benchmarks for FHE, comparing performance across tools is generally difficult
    without implementing a task in a variety of tools.
    Motivated by that, we selected three applications that represent different domains of FHE-based computation.
    Each is designed to showcase complex issues arising when working with FHE, yet also remain simple and easy to reproduce across tools.
    First, we present a risk score calculation that requires comparisons and, therefore, binary circuit emulation.
    While simple, this represents a class of heavily branched programs that is common in traditional programming but hard to express in FHE.
    Second, we consider a statistical $\chi^2$-test, in a formulation that simplifies it to polynomial functions over integers.
    This represents a variety of interesting analysis methods that are ill-suited to FHE by default but can be reformulated or approximated to allow efficient implementations.
        We focus only on the core computation, however in practical deployments, this would probably be preceded by a homomorphic aggregation over user data.
    Finally, we consider machine learning, specifically neural network inference for a range of network architectures. 
    We evaluate a range of increasingly complex models and show how commonly used architectures are adapted for FHE.
    
    In our evaluation, we consider three dimensions: usability, expressiveness, and performance.
    We start by describing each application in detail, 
    then report on the process of implementing these applications in the different tools, highlighting strengths and challenges.
    Where required, we describe adjustments made to the applications due to limits in expressiveness.
    Finally, we present our benchmarking results and highlight the impact of specific techniques or optimizations.

\subsection{Applications}
\subsubsection{Cardio} \label{subsec:cardio}
The cardio risk factor assessment (cardio) application computes a score representing a patient's risk of cardiac disease.
The application takes metrics such as age, gender, weight, drinking habits and smoking behavior where some are integer-valued and others boolean flags as input. 
As illustrated in \Cref{lst:cardio}, the computation consists of a series of simple rules over the inputs that use comparisons and boolean operators.
The algorithm is derived from an implementation in~\cite{Carpov2016-tx}.
Due to its reliance on comparison operations, the program requires emulation using binary circuits.

\begin{lstlisting}[label={lst:cardio},
caption={The Cardio application.},
basicstyle=\ttfamily\small,
escapeinside={(*}{*)}
]
+1  if man && age > 50 years
+1  if woman && age > 60 years
+1  if smoking
+1  if diabetic
+1  if high blood pressure
+1  if HDL cholesterol < 40
+1  if weight > height-90
+1  if daily physical activity < 30
+1  if man && alcohol cons. > 3 glasses/day
+1  if !man && alcohol cons. > 2 glasses/day
\end{lstlisting}

We encode  the inputs as \SI{8}{\bit} numbers and encrypt them at the client-side.
The server receives the encrypted data, computes the risk score, and returns the encrypted score back to the client.
While the inputs are obviously sensitive information, the cardio risk assessment algorithm is public and could easily be calculated client-side.
However,  it is easy to imagine other applications where a service provider might not want to share the algorithm with a client.
For example, similar algorithms are still widely used for risk assessment or fraud detection, and knowledge of the criteria considered makes it easier to circumvent these checks.
For simplicity, we omit the noise flooding required to provide (practical) circuit privacy in our example.

\subsubsection{Chi-Squared Test}

$\chi^2$ or chi-squared tests are common statistical tests.
For our application, we specifically consider Pearson's Goodness-of-Fit test
as it can be used to test for deviation from the Hardy-Weinberg equilibrium in \acf{gwas}.

We split the computation into a polynomial part on the server and a final set of divisions on the client, 
as proposed by Lauter et al.~\cite{Lauter2014-tv}. %
First, the server receives the encrypted genotype counts $N_0, N_1, N_2$, %
then it computes
$ \alpha = (4N_{0}N_{2} - N_{1}^{2})^{2}, \beta_{1} = 2(2N_{0}+N_{1})^{2}, \beta_{2} = (2N_{0}+N_{1})(2N_{2}+N_{1}), \beta_{3} = 2(2N_{2}+N_{1})^{2}$ and returns the encrypted results to the client.
Decrypting these, the client can compute the test statistic as %
$X^{2} = \frac{\alpha}{2N}(\frac{1}{\beta_{1}} + \frac{1}{\beta_{2}} \frac{1}{\beta_{3}}) $.
This transformation introduces some slight leakage of intermediate values but in return enables an application that would otherwise be infeasible.
A more realistic deployment scenario would most likely first see the server calculate the genotype counts over an encrypted genomic database.
While this application is comparatively simple, it is nevertheless practically relevant as seen by its application to genomic studies. 
Additionally, its simplicity allows us to focus more clearly on the overheads introduced by each tool.

\subsubsection{NN Inference}

The neural-network inference application demonstrates FHE's capabilities for privacy-preserving machine learning.
Specifically, we consider inference (or prediction) on a simple image recognition task, i.e., recognizing handwritten digits from the MNIST dataset~\cite{lecun-mnisthandwrittendigit-2010}.
MNIST is a common benchmark in machine learning applications and can be solved effectively by many techniques.
In MNIST, individual inputs are 28\,$\times$\,28 pixel images containing a single handwritten digit.
First, the network is trained over a large number of plaintext images.
Later, a client submits an encrypted input and the model owner returns the encrypted prediction.
This guarantees the privacy of the input and gives strong practical protections for the privacy of the model.
When only the model parameters, but not the general architecture, need to be protected, formal circuit privacy is not required.

\subsection{Implementation Considerations}

In this section, we explain our selection of tools for each application and briefly discuss implementation challenges we faced.
A more detailed documentation of our implementations and design choices is available in our online repository\footnote{\url{https://github.com/MarbleHE/SoK}.}.

\subsubsection{Cardio}

The cardio risk factor assessment requires computing several comparisons between integers, which are hard to approximate polynomially and therefore require binary circuit emulation.
As a baseline, we implemented the programs manually in SEAL and TFHE.
Since EVA targets CKKS, which is less well suited to binary emulation, we do not consider it here.
The Cingulata and \acs{e3} compilers, on the other hand, support binary plaintext spaces natively.

In SEAL and TFHE, we needed to manually implement binary adders and comparators.
This is significantly easier in TFHE, where multiplicative depth is not a concern and a simple ripple-carry-adder is sufficient.
Therefore, our optimized TFHE implementation merely improves the final summation of risk factors by using a tree of adders.
While our naive SEAL implementation also uses a ripple-carry-adder, we also implemented an optimized version where
 we implemented a Sklansky-adder, which trades off additional operations for lower depth.
In the optimized version, we also made heavy use of in-place and plaintext-ciphertext versions of the homomorphic operations, simplified expressions as much as possible, and manually determined optimal parameters.
Finally, we implemented an optimized batched variant, which required significant changes to the computation structure, i.e., transforming all ten conditions into the form \texttt{a \&\& b < c}  by introducing dummy values and operations.

Cingulata makes the implementation significantly more straight-forward as it contains built-in circuits for common operations such as addition, multiplication, and comparisons.
Therefore, the program is virtually identical to its plaintext counterpart. 
However, the compilation process is complex, and the interactions between the compiler and runtime system are not well documented.
This made it hard to integrate the different mult-depth reduction techniques available, and it required significant amounts of trial-and-error to determine, e.g., how Cingulata differentiates between secret and plaintext inputs in the circuits it generates.

\acs{e3} offers a similar and even arguably more powerful API than Cingulata.
For example, it supports both binary and arithmetic plaintext spaces and can switch ciphertexts between them.
In a similar vein, very few changes were needed to re-target our SEAL (BFV) implementation to TFHE (CGGI).
However, an initial lack of documentation and very long compile times made developing and debugging applications difficult.
While \ac{e3} features some support for batching, this is quite limited.
Specifically, it does not include rotation operations that are essential to fully express the program's batched version.
Therefore, the \ac{e3} batched version remains somewhat incomplete.

\subsubsection{Chi-Squared Test}
The Chi-Squared test, at least as re-formulated in our application, uses only addition and multiplication over integers, making it ideally suited for integer-based FHE schemes. 
Nevertheless, we also consider implementations targeting binary emulation for comparison.
We manually implemented the application in SEAL, targeting the BFV schemes and an integer plaintext space.
In our optimized version, we manually select optimal parameters, use in-place operations where possible and reuse common sub-expressions.
Our manual implementations in SEAL closely match the mathematical description as all operations are native operations.
Nevertheless, both the naive and the optimized implementation required over 100 lines of code.
Our TFHE-based manual implementations additionally required implementing a binary adder and multiplier to support the computation, resulting in several hundred lines of code.
EVA, in contrast, allowed us to easily express the same computation in around a dozen lines of code.
While the EVA implementation targets CKKS, the precision is sufficient to ensure that, when rounding back to integers, the result perfectly matches the other BFV/integer-based implementations.
While Cingulata supports the BFV scheme, it only supports binary plaintext spaces.
Therefore, it must also emulate integer multiplications using binary circuits.
However, since it hides the complexity of generating efficient circuits from the user, this matters only for performance, not for usability.
Both Cingulata and \ac{e3} can target integer-based BFV and binary CGGI with minimal changes required.
Note that batching this application would be trivial but only impacts throughput, not latency, and is therefore omitted.

\subsubsection{NN-Inference} 

The MNIST problem is comparatively easy to solve, with simple approaches easily achieving more than $90\%$ accuracy and even small neural networks achieving around $95\%$ accuracy.
State-of-the-art networks achieve up to $99.5\%$ test accuracy.
However, increasing accuracy quickly requires exponentially more complex models.
In our evaluation, we used three different model architectures of increasing complexity.
First, we used a simple Multi-Layer Perceptron~(MLP) as a baseline, i.e., two fully connected layers with a non-linear activation.
Next, we consider a more complicated Convolutional Neural Network~(CNN), specifically the Cryptonets architecture~\cite{Gilad-Bachrach2016-aq} designed specifically for FHE, 
which consists of 5 layers and two activations.
Finally, we also evaluated a LeNet-5-like~\cite{lecun1998gradient} network, which is a significantly more complex design and more representative of networks used to solve challenging tasks in practice.
This network consists of 7 layers and three activations.
We use a technique from~\cite{dathathri2018chet} and learn a degree-two polynomial approximation of the ReLU activation function during training.

SEALion and nGraph-HE focus exclusively on machine learning inference, directly using TensorFlow programs or TensorFlow-like programs as their inputs.
While SEALion can currently only express a simple MLP network, nGraph-HE seems to support the full TensorFlow feature set.
Both make FHE-based development nearly as easy as working with standard TensorFlow.
While EVA does not directly support machine learning tasks, the CHET tool can be re-targeted to EVA, and we consider an EVA program for a LeNet-5 model generated by CHET, in addition to a manually implemented MLP.
We complemented the comparison between the tools with a baseline implementation of an MLP in SEAL, using the CKKS scheme and manually implementing matrix-vector-product optimizations from~\cite{juvekar2018}, which required significant engineering effort.

\subsection{Effects of Optimizations} \label{subsec:effects-of-optimizations}
This section presents the results of our benchmarks, with a particular focus on the effect that automation and optimization have on runtime.
All benchmarks run on an AWS instance (m5n.xlarge), equipped with 4 vCPUs and \SI{16}{\giga\byte} RAM.
The reported results are mean values computed over 10 test runs.

\subsubsection{Cardio}

\begin{figure}[tbp]
    \centering
    \includegraphics[width=\columnwidth]{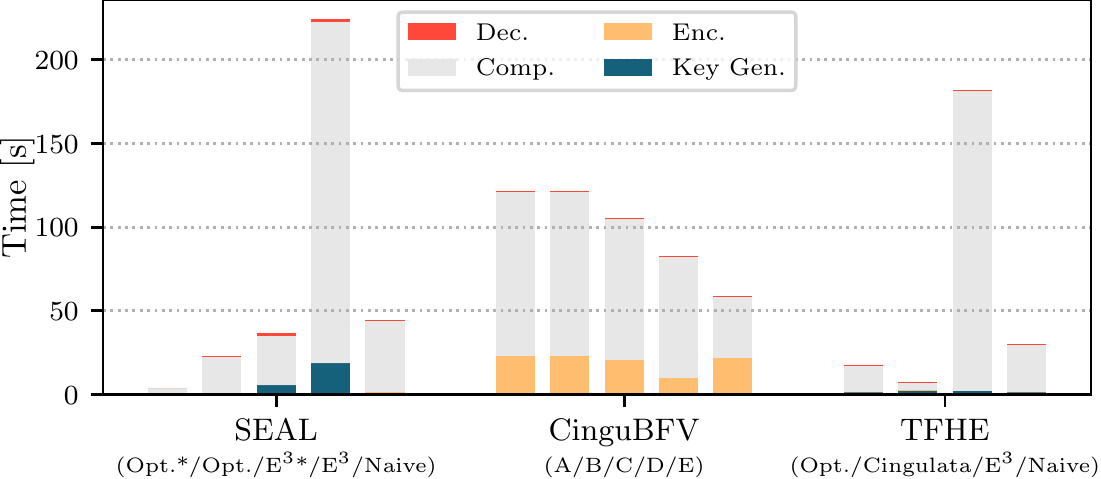}
    \caption{Runtime of the cardio benchmark.
        We group compiler generated and manually optimized and naive programs by the FHE implementation they target.        
        For CinguBFV, we consider circuits using different depth-optimization approaches (A: baseline, B: ABC, C: Lobster, D: Cingulata, E: Multi-Start).
        * indicates batching was used. 
    }
    \label{fig:plot-cardio}
    \vspace{-5pt}
\end{figure}

In \Cref{fig:plot-cardio}, we report the run time for the cardio risk factor assessment application in different setups.
We see a large span of results, between less than 5~seconds  for the manual optimized implementation 
and over three minutes for the slowest tool-generated implementations.
\ac{e3} seems to introduce significant overheads, even when compared to naive implementations targeting the same library.
Cingulata's BFV implementation (CinguBFV) seems considerably slower than SEAL's, but we can still observe the effect of the different depth-reduction approaches, with multi-start (E) cutting computation time in half.
Comparing our manual implementations, we see both of our TFHE implementations outperforming the naive and (non-batched) optimized SEAL implementation as expected.
    Cingulata's TFHE implementation actually further outperforms our manually optimized TFHE implementation, even when our manual program uses fewer gates.
    This speedup might be due to better memory management or due to slightly different TFHE environments.
However, by far the best performance is achieved when using batching in SEAL, even though this application is inherently binary-based and ten conditions are a relatively small number to batch in the context of FHE.

\subsubsection{Chi-Squared Test}

\begin{figure}[tbp]
    \centering
    \includegraphics[width=\columnwidth]{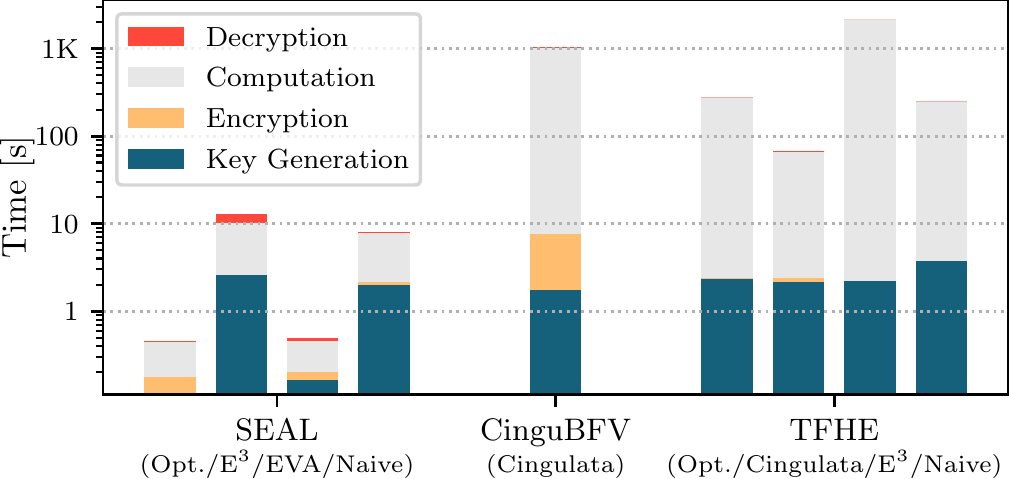}
    \caption{Runtime of the chi-squared test benchmark using a logarithmic scale. 
        We group compiler generated, optimized and naive programs by the FHE implementation they target.      
}
    \label{fig:plot-chi-squared}
    \vspace{-10pt}
\end{figure}

In \Cref{fig:plot-chi-squared}, we present the runtimes for the chi-squared test application,
using a logarithmic scale due to the large range of values.
We contrast manually- and tool-generated implementations targeting SEAL and TFHE and compare this against Cingulata's implementation targeting the built-in BFV implementation. 
The manually optimized SEAL implementation and EVA-generated implementation outperform the others by a large margin, requiring less than a second. %
With \SI{16.46}{\s}, a slowdown of more than 10$\times$, the E3 program targeting SEAL is significantly slower,
but the overhead compared to the naive solution is negligible.
Meanwhile, Cingulata targeting CinguBFV suffers from both using binary emulation unnecessarily and a generally slower BFV implementation.
    Since the program already has minimal depth, we omit a discussion of the different depth-reduction heuristics here.
    Similarly, our TFHE optimizations seem to have no positive effect on this simple program, 
    while Cingulata is again faster per-gate in TFHE, possibly due to configuration differences.
Finally, we note that the TFHE implementation generated by E3 is around 9$\times$ slower than native implementations, 
which are already non-competitive compared to integer-based solutions.
In combination with the cardio benchmarking results, this indicates that \ac{e3} generates binary adder/multiplier circuits inefficiently when using binary emulation.

\subsubsection{NN Inference}
\begin{figure}[tbp]
    \centering
    \includegraphics[width=\columnwidth]{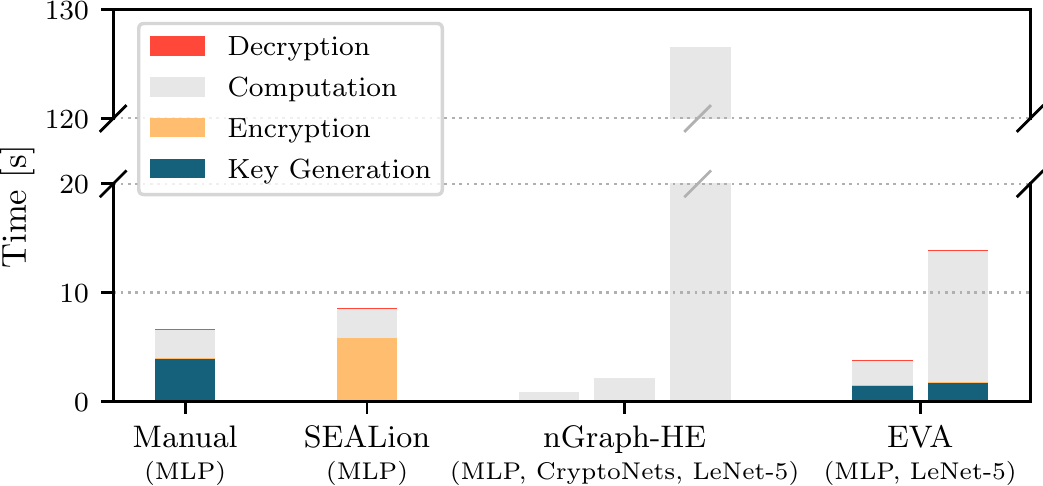}
    \caption{Runtime of the neural network inference benchmark, i.e., recognizing handwritten digits from the MNIST dataset. 
        All implementations target SEAL.
        We implement a simple multi-layer-perceptron (MLP). For nGraph-HE and EVA, we also consider more complex models (CryptoNets, LeNet-5).
    }
    \label{fig:plot-nn}
    \vspace{-10pt}
\end{figure}

We present the evaluation results for the neural-network inference task in \Cref{fig:plot-nn}, 
reporting latency, i.e., the time to run encrypted prediction on a single image.
Note that SEALion and nGraph-HE use \ac{simd}-style batching to achieve higher throughput at the same latency.
For nGraph-HE, it was not possible to provide individual sub-timings, as key-generation, encryption, and decryption are invisible to the application code. %
We first compare our manual implementation of an MLP both directly in SEAL and using EVA against the same network architecture implemented in SEALion and nGraph-HE, which offer much higher-level interfaces.
All models achieved around 95\% accuracy, nearly identical to their plaintext equivalents.
Note that for SEALion, the overall runtime is artificially inflated because the tool encrypts the input against a range of possible parameter sets instead of only the targeted one.
Taking this into account, we can see that despite us implementing several optimization techniques from the literature,
the higher-level tools clearly outperform the manual implementation.
In the case of SEALion, this appears to be due to automatic sparsification, which reduces the network's size. %
Finally, we explored more complex models using nGraph-HE and EVA, using CHET-generated programs for the latter.
The Cryptonets CNN architecture significantly increases accuracy (to 98\%) at a minimal increase in computation cost.
However, achieving state-of-the-art network performance (99+\%) requires a considerably more complex LeNet-5-like network, which takes around 13~seconds to run using EVA and more than two minutes using nGraph-HE.

\section{Discussion}
In this section, we discuss some key questions in the space of FHE and FHE tools:

    \subsection{What applications can be developed using FHE today?} %
    While FHE can be practical for a wide variety of applications, there remain many applications that are not yet feasible using FHE.
    Applications that make sense for FHE generally feature a client-server scenario where both the input data and the algorithm need to be kept private.
    In addition, there are practical limits to the complexity of the applications that can be outsourced.
    As a very rough heuristic, computations that take more than a few hundred milliseconds without FHE are unlikely to be practical once translated into FHE as of today.
    However, this very much depends on the application scenario.
    Generally, online computations where immediate feedback is expected are more challenging.
    For example, face recognition applications at an airport might tolerate a few seconds of delay at most.
    On the other hand, offline tasks like computing statistics over the results of a year-long medical study can be considered practical even if taking considerable time.
     
    For non-expert users, the range of applications that can be realized in practice also depends significantly on the available tools.
     Using libraries like \acs{seal}, \Palisade, or \acs{helib} makes it easy to implement simple computations that can be expressed as low-degree polynomials (e.g., the modified $\chi^2$ test), and tools like nGraph-HE enable novice users to easily implement linear ML models, simple statistics, and neural network inference.
    For more experienced users, this question becomes increasingly difficult to address in general terms.
    The implementation challenges we describe for our case studies show that application complexity and FHE implementation complexity do not necessarily correlate.
    Finally, some applications require modifications or extensions of the underlying cryptographic primitives.
    These include computations that require switching between different schemes or between FHE and MPC homomorphically.
      Many applications can already be solved practically using these or other novel programming paradigms.
      Frequently, success in implementing an efficient FHE-based solution for an application depends less on the performance of the underlying FHE tools but on how the application is translated.
      Exploiting the advantage of SIMD-batching, e.g., using EVA, requires designing heavily vectorized programs
      for a setting with significantly more restrictions than, e.g., AVX vector instructions.
      In addition, many applications become feasible only after slight modifications,e.g.,
      using polynomial approximations or rewriting expressions so that hard-to-compute operations (e.g., square roots) are delayed until the end to allow them to be performed client-side after decryption.        
    By presenting these paradigms more clearly and targeting an audience beyond the crypto community,  
    the set of applications that developers can expect to realize successfully using FHE will expand significantly.
	       
    \subsection{When to use which of the FHE tools?}

        Given the choice of different tools that each present slightly different features and strengths,
        selecting the appropriate tool for a given application can be non-trivial.
        However, not all tools that \emph{can} implement a solution are necessarily suitable choices, as demonstrated in our evaluation.
        Current tools generally excel at specific workloads or application domains, 
        and here we try to provide some recommendations for tools to consider for common application scenarios.

    For generic applications that compute non-polynomial functions or require binary emulation,
    there are multiple options with different tradeoffs.
    If working primarily with integers, the programmable bootstrapping offered by the \emph{concrete library} is an obvious choice.
    While compilers like Cingulata (CinguBFV) or \acs{e3} are easier to work with, the performance overhead they introduce might be unacceptable for many applications.
    For applications requiring a true binary plaintext space, \emph{Cingulata (TFHE)} is most likely the easiest approach.

    For applications that compute (polynomial) statistics over large amounts of data,
    we recommend the \emph{EVA compiler} targeting \ac{ckks} for applications requiring approximate numbers.
    If working with integers only, we recommend working directly with the \emph{SEAL library} targeting \ac{bfv},
     since BFV is less complex to work with and current compilers targeting it introduce significant slowdowns.
    The batching offered by these schemes can be a natural fit when computing aggregate statistics or retrieving information from encrypted databases.

     For applications that involve or use machine learning inference, the recommended approach depends on the complexity of the used ML model.
     Where training a model with polynomial activation functions produces sufficient accuracy, 
     we recommend using the \emph{nGraph-HE compiler} targeting the \ac{ckks} scheme.
      nGraph-HE supports virtually all TensorFlow features, including the Keras model definition API, making it trivial to port existing models.
      In addition, nGraph-HE offers excellent performance that can easily outperform even a fairly involved manual implementation. 
        Where deeper/recursive networks or standard activation functions (e.g., ReLU) are required to achieve the desired accuracy, 
          the programmable bootstrapping functionality offered by the \emph{concrete library} makes it the most suitable choice.
          However, this will require significantly more engineering effort as there are currently no higher-level compilers targeting concrete.

\vspace{-2pt}
     
    \subsection{Where should FHE tools go from here?}
    Both FHE compilers and libraries remain complex to use, and there are obvious low-hanging fruits in terms of usability that include better documentation and more extensive examples.
    In addition, there is a general lack of interoperability, not just technically but also conceptually.
    For example, even libraries implementing the same scheme can offer surprisingly different APIs.
    The ongoing standardization efforts are trying to create a unified view of the most popular schemes, including standardized APIs for the most common operations.
  However, this does not address the various extension of the API, e.g., optimizations for squaring rather than multiplying or performing operations in-place.
  This would be solved ideally by introducing a common intermediate representation language that compilers can target and libraries can implement.

         The existing tools have successfully reduced the complexity of working with complex FHE schemes.
         There is a large choice of libraries providing secure and efficient implementations of current schemes.
         In addition, compilers have emerged that make it significantly easier to realize computations efficiently, e.g., by automatically choosing parameters or inserting ciphertext maintenance operations.
         However, this still leaves the user with the significant challenge of translating an application into an appropriate FHE computation in the first place.
         For example, tools could automatically vectorize iteratively written programs or offer suggestions on aspects of the computation that would be beneficial to extract to the client-side.

  Finally, it is worth noting that we have considered only FHE tools in our analysis and discussion.
  However, real-world problems are frequently complex and require a combination of techniques, including FHE, \acf{mpc}, and Zero-Knowledge Proofs (ZKP).
  In the long term, the secure computation community could gain tremendously by considering these problems more holistically and building tools that support a wider range of techniques.

\section*{Acknowledgments}
We thank our shepherd, the anonymous reviewers, and Kenny Paterson for their valuable feedback. 
We thank the FHE tool developers and
maintainers for making their code available and the CHET/EVA and SEALion developers for their help and for providing us with access to their tools.
This work was supported in part by the SNSF Ambizione Grant No.~186050.

\bibliographystyle{IEEEtran}
\bibliography{IEEEabrv,references}

\begin{thebibliography}{10}
\providecommand{\url}[1]{#1}
\csname url@samestyle\endcsname
\providecommand{\newblock}{\relax}
\providecommand{\bibinfo}[2]{#2}
\providecommand{\BIBentrySTDinterwordspacing}{\spaceskip=0pt\relax}
\providecommand{\BIBentryALTinterwordstretchfactor}{4}
\providecommand{\BIBentryALTinterwordspacing}{\spaceskip=\fontdimen2\font plus
\BIBentryALTinterwordstretchfactor\fontdimen3\font minus
  \fontdimen4\font\relax}
\providecommand{\BIBforeignlanguage}[2]{{%
\expandafter\ifx\csname l@#1\endcsname\relax
\typeout{** WARNING: IEEEtran.bst: No hyphenation pattern has been}%
\typeout{** loaded for the language `#1'. Using the pattern for}%
\typeout{** the default language instead.}%
\else
\language=\csname l@#1\endcsname
\fi
#2}}
\providecommand{\BIBdecl}{\relax}
\BIBdecl

\bibitem{saleem2020sok}
\BIBentryALTinterwordspacing
H.~Saleem and M.~Naveed, ``{SoK: Anatomy of Data Breaches},'' \emph{Proceedings
  on Privacy Enhancing Technologies}, vol. 2020, no.~4, pp. 153--174, 2020.
  [Online]. Available: \url{http://isyou.info/jowua/papers/jowua-v10n4-4.pdf}
\BIBentrySTDinterwordspacing

\bibitem{enterprise-data-breach}
L.~Cheng, F.~Liu, and D.~D. Yao, ``{Enterprise Data Breach: Causes, Challenges,
  Prevention, and future Directions},'' \emph{{Wiley Interdisciplinary Reviews:
  Data Mining and Knowledge Discovery}}, vol.~7, no.~5, 2017.

\bibitem{Zama}
\BIBentryALTinterwordspacing
``Zama,'' accessed: 2020-12-21. [Online]. Available: \url{https://zama.ai/}
\BIBentrySTDinterwordspacing

\bibitem{Enveil2020-ds}
\BIBentryALTinterwordspacing
{Enveil}, ``Enveil raises \$10 million in series a funding,'' 18~Feb. 2020,
  accessed: 2020-12-21. [Online]. Available:
  \url{https://www.globenewswire.com/news-release/2020/02/18/1986152/0/en/Enveil-Raises-10-Million-in-Series-A-Funding.html}
\BIBentrySTDinterwordspacing

\bibitem{Inpher2018-ry}
\BIBentryALTinterwordspacing
{Inpher}, ``{J.P}. morgan leads {USD} \$10 million financing in leading data
  security and machine learning provider, inpher,'' 2~Nov. 2018, accessed:
  2020-12-21. [Online]. Available:
  \url{https://www.prnewswire.com/news-releases/jp-morgan-leads-usd-10-million-financing-in-leading-data-security-and-machine-learning-provider-inpher-300743090.html}
\BIBentrySTDinterwordspacing

\bibitem{Lunden2019-nx}
\BIBentryALTinterwordspacing
I.~Lunden, ``Duality, a security startup co-founded by the creator of
  homomorphic encryption, raises \$16m,'' \emph{TechCrunch}, 30~Oct. 2019.
  [Online]. Available:
  \url{http://techcrunch.com/2019/10/30/duality-cybersecurity-16-million/}
\BIBentrySTDinterwordspacing

\bibitem{Jain_undated-dh}
\BIBentryALTinterwordspacing
R.~Jain, ``Data encryption provider {IXUP} appoints new {CEO} \& {MD} marcus
  gracey,'' accessed: 2020-12-21. [Online]. Available:
  \url{https://itmunch.com/data-encryption-provider-ixup-appoints-new-ceo-md-marcus-gracey/}
\BIBentrySTDinterwordspacing

\bibitem{Loritz_undated-aq}
\BIBentryALTinterwordspacing
M.~Loritz, ``Paris-based cosmian raises €1.4 for its platform that analyses
  encrypted data while keeping it private,'' accessed: 2020-12-21. [Online].
  Available:
  \url{https://www.eu-startups.com/2019/03/paris-based-cosmian-raises-e1-4-for-its-platform-that-analyses-encrypted-data-while-keeping-it-private/}
\BIBentrySTDinterwordspacing

\bibitem{Osborne2020-yj}
\BIBentryALTinterwordspacing
C.~Osborne, ``{IBM} launches experimental homomorphic data encryption
  environment for the enterprise,'' Dec. 2020, accessed: 2020-12-21. [Online].
  Available:
  \url{https://www.zdnet.com/article/ibm-launches-experimental-homomorphic-data-encryption-environment-for-the-enterprise/}
\BIBentrySTDinterwordspacing

\bibitem{Rivest1978a}
\BIBentryALTinterwordspacing
R.~L. Rivest, L.~Adleman, and M.~L. Dertouzos, ``On data banks and privacy
  homomorphisms,'' \emph{Foundations of secure computation}, vol.~4, no.~11,
  pp. 169--180, 1978. [Online]. Available:
  \url{https://people.csail.mit.edu/rivest/RivestAdlemanDertouzos-OnDataBanksAndPrivacyHomomorphisms.pdf}
\BIBentrySTDinterwordspacing

\bibitem{Gentry2009-zi}
\BIBentryALTinterwordspacing
C.~Gentry, ``A fully homomorphic encryption scheme,'' Ph.D. dissertation,
  Stanford University, 2009. [Online]. Available:
  \url{https://crypto.stanford.edu/craig/}
\BIBentrySTDinterwordspacing

\bibitem{Smart2014}
\BIBentryALTinterwordspacing
N.~P. Smart and F.~Vercauteren, ``Fully homomorphic {{SIMD}} operations,''
  \emph{Designs, Codes and Cryptography. An International Journal}, vol.~71,
  no.~1, pp. 57--81, Jan. 2014. [Online]. Available:
  \url{https://doi.org/10.1007/s10623-012-9720-4}
\BIBentrySTDinterwordspacing

\bibitem{Microsoft2019-qj}
\BIBentryALTinterwordspacing
{Microsoft}, ``{AsureRun},'' 11~May 2019. [Online]. Available:
  \url{https://github.com/microsoft/SEAL-Demo/tree/master/AsureRun}
\BIBentrySTDinterwordspacing

\bibitem{Kim2020-dk}
\BIBentryALTinterwordspacing
M.~Kim, A.~Harmanci, J.-P. Bossuat, S.~Carpov, J.~H. Cheon, I.~Chillotti,
  W.~Cho, D.~Froelicher, N.~Gama, M.~Georgieva, S.~Hong, J.-P. Hubaux, D.~Kim,
  K.~Lauter, Y.~Ma, L.~{Ohno-Machado}, H.~Sofia, Y.~Son, Y.~Song,
  J.~{Troncoso-Pastoriza}, and X.~Jiang,
  ``\BIBforeignlanguage{English}{Ultra-{{Fast}} homomorphic encryption models
  enable secure outsourcing of genotype imputation},'' May 2020. [Online].
  Available: \url{https://www.biorxiv.org/content/10.1101/ 2020.07.02.183459v2}
\BIBentrySTDinterwordspacing

\bibitem{Chen2018a}
\BIBentryALTinterwordspacing
H.~Chen, Z.~Huang, K.~Laine, and P.~Rindal, ``Labeled {{PSI}} from {{Fully
  Homomorphic Encryption}} with {{Malicious Security}},'' in \emph{Proceedings
  of the 2018 {{ACM SIGSAC Conference}} on {{Computer}} and {{Communications
  Security}}}.\hskip 1em plus 0.5em minus 0.4em\relax {Toronto Canada}: {ACM},
  Jan. 2018, pp. 1223--1237. [Online]. Available:
  \url{https://eprint.iacr.org/2018/787}
\BIBentrySTDinterwordspacing

\bibitem{Kim2019}
\BIBentryALTinterwordspacing
M.~Kim, Y.~Song, B.~Li, and D.~Micciancio, ``Semi-{{Parallel Logistic
  Regression}} for {{GWAS}} on {{Encrypted Data}}.'' \emph{IACR Cryptology
  ePrint Archive}, vol. 2019, p. 294, 2019. [Online]. Available:
  \url{https://eprint.iacr.org/2019/294}
\BIBentrySTDinterwordspacing

\bibitem{Dathathri2019-vu}
\BIBentryALTinterwordspacing
R.~Dathathri, B.~Kostova, O.~Saarikivi, W.~Dai, K.~Laine, and M.~Musuvathi,
  ``{EVA}: An encrypted vector arithmetic language and compiler for efficient
  homomorphic computation,'' in \emph{Proceedings of the 41st {ACM} {SIGPLAN}
  Conference on Programming Language Design and Implementation}, 27~Dec. 2019.
  [Online]. Available: \url{http://arxiv.org/abs/1912.11951}
\BIBentrySTDinterwordspacing

\bibitem{chou2020}
E.~J. Chou, A.~Gururajan, K.~Laine, N.~K. Goel, A.~Bertiger, and J.~W. Stokes,
  ``Privacy-preserving phishing web page classification via fully homomorphic
  encryption,'' in \emph{{{ICASSP}} 2020 - 2020 {{IEEE}} International
  Conference on Acoustics, Speech and Signal Processing ({{ICASSP}})}, 2020,
  pp. 2792--2796.

\bibitem{Driver2020-gc}
M.~Driver, ``Emerging technologies: Homomorphic encryption for data sharing
  with privacy,'' Gartner, Inc, Tech. Rep., 23~Apr. 2020.

\bibitem{sealcrypto}
\BIBentryALTinterwordspacing
``Microsoft {{SEAL}} (release 3.5),'' Apr. 2020. [Online]. Available:
  \url{https://github.com/Microsoft/SEAL}
\BIBentrySTDinterwordspacing

\bibitem{TFHE}
\BIBentryALTinterwordspacing
I.~Chillotti, N.~Gama, M.~Georgieva, and M.~Izabach{\`e}ne, ``{{TFHE}}:
  {{Fast}} fully homomorphic encryption library,'' Aug. 2016. [Online].
  Available: \url{https://tfhe.github.io/tfhe}
\BIBentrySTDinterwordspacing

\bibitem{HomomorphicEncryptionSecurityStandard}
\BIBentryALTinterwordspacing
M.~Albrecht, M.~Chase, H.~Chen, J.~Ding, S.~Goldwasser, S.~Gorbunov, S.~Halevi,
  J.~Hoffstein, K.~Laine, K.~Lauter, S.~Lokam, D.~Micciancio, D.~Moody,
  T.~Morrison, A.~Sahai, and V.~Vaikuntanathan, ``Homomorphic encryption
  security standard,'' {HomomorphicEncryption.org}, {Toronto, Canada}, Tech.
  Rep., Nov. 2018. [Online]. Available: \url{https://homomorphicencryption.org}
\BIBentrySTDinterwordspacing

\bibitem{Barlow2019-si}
\BIBentryALTinterwordspacing
N.~Barlow, T.~Lazauskas, O.~Strickson, and A.~Gascon, ``{{SHEEP}}: A
  homomorphic encryption evaluation platform,'' Nov. 2019. [Online]. Available:
  \url{https://github.com/alan-turing-institute/SHEEP}
\BIBentrySTDinterwordspacing

\bibitem{Boemer2019-mt}
\BIBentryALTinterwordspacing
F.~Boemer, A.~Costache, R.~Cammarota, and C.~Wierzynski, ``{{nGraph}}-{{HE2}}:
  {{A High}}-{{Throughput}} framework for neural network inference on encrypted
  data,'' in \emph{Proceedings of the 7th {{ACM}} Workshop on Encrypted
  Computing \& Applied Homomorphic Cryptography}, ser. {{WAHC}}'19.\hskip 1em
  plus 0.5em minus 0.4em\relax {New York, NY, USA}: {Association for Computing
  Machinery}, Nov. 2019, pp. 45--56. [Online]. Available:
  \url{https://doi.org/10.1145/3338469.3358944}
\BIBentrySTDinterwordspacing

\bibitem{Acar2018}
A.~Acar, H.~Aksu, A.~S. Uluagac, and M.~Conti, ``A {{Survey}} on {{Homomorphic
  Encryption Schemes}}: {{Theory}} and {{Implementation}},'' \emph{ACM
  Computing Surveys}, vol.~51, no.~4, pp. 1--35, Jul. 2018.

\bibitem{Hastings2019-qw}
\BIBentryALTinterwordspacing
M.~Hastings, B.~Hemenway, D.~Noble, and S.~Zdancewic,
  ``\BIBforeignlanguage{English}{{{SoK}}: {{General}} purpose compilers for
  secure {{Multi}}-{{Party}} computation},'' in
  \emph{\BIBforeignlanguage{English}{{{IEEE}} Symposium on Security and Privacy
  ({{SP}})}}.\hskip 1em plus 0.5em minus 0.4em\relax {Los Alamitos, CA, USA}:
  {IEEE Computer Society}, 2019, pp. 479--496. [Online]. Available:
  \url{https://www.computer.org/csdl/proceedings/sp/2019/6660/00/
  666000a462-abs.html}
\BIBentrySTDinterwordspacing

\bibitem{Paillier1999a}
\BIBentryALTinterwordspacing
P.~Paillier, ``\BIBforeignlanguage{English}{Public-{{Key}} cryptosystems based
  on composite degree residuosity classes},'' in
  \emph{\BIBforeignlanguage{English}{Advances in Cryptology \textemdash{}
  {{EUROCRYPT}} '99}}, ser. {{EUROCRYPT}}.\hskip 1em plus 0.5em minus
  0.4em\relax {Prague, Czech Republic}: {Springer, Berlin, Heidelberg}, May
  1999, pp. 223--238. [Online]. Available:
  \url{http://link.springer.com/chapter/10.1007/3-540-48910-X_16}
\BIBentrySTDinterwordspacing

\bibitem{Rivest1978}
\BIBentryALTinterwordspacing
R.~L. Rivest, A.~Shamir, and L.~Adleman, ``A method for obtaining digital
  signatures and public-key cryptosystems,'' \emph{Communications of the ACM},
  vol.~21, no.~2, pp. 120--126, Feb. 1978. [Online]. Available:
  \url{http://dl.acm.org/citation.cfm?id=359342\&}
\BIBentrySTDinterwordspacing

\bibitem{Boneh2005}
\BIBentryALTinterwordspacing
D.~Boneh, E.-J. Goh, and K.~Nissim, ``\BIBforeignlanguage{English}{Evaluating
  2-{{DNF}} formulas on ciphertexts},'' in
  \emph{\BIBforeignlanguage{English}{Theory of Cryptography}}.\hskip 1em plus
  0.5em minus 0.4em\relax {Springer, Berlin, Heidelberg}, Feb. 2005, pp.
  325--341. [Online]. Available:
  \url{http://link.springer.com/chapter/10.1007/978-3-540-30576-7_18}
\BIBentrySTDinterwordspacing

\bibitem{Gentry2011-kk}
C.~Gentry and S.~Halevi, ``Implementing gentry's {{Fully}}-{{Homomorphic}}
  encryption scheme,'' in \emph{{{EUROCRYPT}}}, 2011.

\bibitem{Brakerski2014-uq}
\BIBentryALTinterwordspacing
Z.~Brakerski, C.~Gentry, and V.~Vaikuntanathan, ``(leveled) fully homomorphic
  encryption without bootstrapping,'' \emph{ACM Transactions on Computation
  Theory}, vol.~6, no.~3, pp. 13:1--13:36, Jul. 2014. [Online]. Available:
  \url{http://doi.acm.org/10.1145/2633600}
\BIBentrySTDinterwordspacing

\bibitem{fan2012somewhat}
\BIBentryALTinterwordspacing
J.~Fan and F.~Vercauteren, ``Somewhat {{Practical Fully Homomorphic
  Encryption}},'' \emph{IACR Cryptology ePrint Archive}, vol. 2012, p. 144,
  2012. [Online]. Available: \url{https://eprint.iacr.org/2012/144}
\BIBentrySTDinterwordspacing

\bibitem{brakerski2012fully}
\BIBentryALTinterwordspacing
Z.~Brakerski, ``Fully {{Homomorphic Encryption}} without {{Modulus Switching}}
  from {{Classical GapSVP}},'' in \emph{Advances in {{Cryptology}}
  \textendash{} {{CRYPTO}} 2012}.\hskip 1em plus 0.5em minus 0.4em\relax
  {Berlin, Heidelberg}: {Springer Berlin Heidelberg}, 2012, vol. 7417, pp.
  868--886. [Online]. Available:
  \url{https://link.springer.com/chapter/10.1007/978-3-642-32009-5_50}
\BIBentrySTDinterwordspacing

\bibitem{iliashenko2019optimisations}
\BIBentryALTinterwordspacing
I.~Iliashenko, ``Optimisations of fully homomorphic encryption,'' Ph.D.
  dissertation, PhD thesis, KU Leuven, 2019. [Online]. Available:
  \url{https://www.esat.kuleuven.be/cosic/publications/thesis-316.pdf}
\BIBentrySTDinterwordspacing

\bibitem{halevi2018faster}
\BIBentryALTinterwordspacing
S.~Halevi and V.~Shoup, ``Faster {{Homomorphic Linear Transformations}} in
  {{HElib}},'' in \emph{Advances in {{Cryptology}} \textendash{} {{CRYPTO}}
  2018}.\hskip 1em plus 0.5em minus 0.4em\relax {Cham}: {Springer International
  Publishing}, 2018, vol. 10991, pp. 93--120. [Online]. Available:
  \url{https://link.springer.com/chapter/10.1007/978-3-319-96884-1_4}
\BIBentrySTDinterwordspacing

\bibitem{cheon2017homomorphic}
\BIBentryALTinterwordspacing
J.~H. Cheon, A.~Kim, M.~Kim, and Y.~Song, ``Homomorphic {{Encryption}} for
  {{Arithmetic}} of {{Approximate Numbers}},'' in \emph{Advances in
  {{Cryptology}} \textendash{} {{ASIACRYPT}} 2017}.\hskip 1em plus 0.5em minus
  0.4em\relax {Cham}: {Springer International Publishing}, 2017, vol. 10624,
  pp. 409--437. [Online]. Available:
  \url{https://www.springerprofessional.de/homomorphic-encryption-for-arithmetic-of-approximate-numbers/15266370}
\BIBentrySTDinterwordspacing

\bibitem{gentry2013homomorphic}
C.~Gentry, A.~Sahai, and B.~Waters, ``Homomorphic {{Encryption}} from
  {{Learning}} with {{Errors}}: {{Conceptually}}-{{Simpler}},
  {{Asymptotically}}-{{Faster}}, {{Attribute}}-{{Based}},'' in \emph{Advances
  in {{Cryptology}} \textendash{} {{CRYPTO}} 2013}.\hskip 1em plus 0.5em minus
  0.4em\relax {Berlin, Heidelberg}: {Springer Berlin Heidelberg}, 2013, vol.
  8042, pp. 75--92.

\bibitem{Chillotti2016Dec}
\BIBentryALTinterwordspacing
I.~Chillotti, N.~Gama, M.~Georgieva, and M.~Izabach{\`e}ne, ``Faster {{Fully
  Homomorphic Encryption}}: {{Bootstrapping}} in {{Less Than}} 0.1
  {{Seconds}},'' \emph{SpringerLink}, pp. 3--33, Dec. 2016. [Online].
  Available: \url{https://eprint.iacr.org/2016/870}
\BIBentrySTDinterwordspacing

\bibitem{Chillotti2017-yd}
\BIBentryALTinterwordspacing
------, ``Faster packed homomorphic operations and efficient circuit
  bootstrapping for {{TFHE}},'' in \emph{Advances in Cryptology \textendash{}
  {{ASIACRYPT}} 2017}.\hskip 1em plus 0.5em minus 0.4em\relax {Springer
  International Publishing}, 2017, pp. 377--408. [Online]. Available:
  \url{https://link.springer.com/chapter/10.1007/978-3-319-70694-8_14}
\BIBentrySTDinterwordspacing

\bibitem{Chillotti2020-ia}
I.~Chillotti, M.~Joye, and P.~Paillier, ``Programmable bootstrapping enables
  efficient homomorphic inference of deep neural networks,'' Zama, Tech. Rep.,
  15~Oct. 2020.

\bibitem{Mouchet2020-pz}
\BIBentryALTinterwordspacing
C.~Mouchet, J.~{Troncoso-Pastoriza}, and J.-P. Hubaux, ``Multiparty homomorphic
  encryption: {{From}} theory to practice,'' 2020. [Online]. Available:
  \url{https://eprint.iacr.org/2020/304}
\BIBentrySTDinterwordspacing

\bibitem{Chen2019-ja}
\BIBentryALTinterwordspacing
H.~Chen, W.~Dai, M.~Kim, and Y.~Song, ``Efficient {{Multi}}-{{Key}} homomorphic
  encryption with packed ciphertexts with application to oblivious neural
  network inference,'' in \emph{Proceedings of the 2019 {{ACM SIGSAC}}
  Conference on Computer and Communications Security}, ser. {{CCS}} '19.\hskip
  1em plus 0.5em minus 0.4em\relax {New York, NY, USA}: {Association for
  Computing Machinery}, Jun. 2019, pp. 395--412. [Online]. Available:
  \url{https://doi.org/10.1145/3319535.3363207}
\BIBentrySTDinterwordspacing

\bibitem{juvekar2018}
\BIBentryALTinterwordspacing
C.~Juvekar, V.~Vaikuntanathan, and A.~Chandrakasan, ``{{GAZELLE}}: {{A Low
  Latency Framework}} for {{Secure Neural Network Inference}},'' in
  \emph{Proceedings of the 27th {{USENIX}} Conference on Security Symposium},
  ser. {{SEC}}'18.\hskip 1em plus 0.5em minus 0.4em\relax {Berkeley, CA, USA}:
  {USENIX Association}, 2018, pp. 1651--1668. [Online]. Available:
  \url{https://www.usenix.org/conference/usenixsecurity18/presentation/juvekar}
\BIBentrySTDinterwordspacing

\bibitem{Boura2018-wj}
\BIBentryALTinterwordspacing
C.~Boura, N.~Gama, and M.~Georgieva, ``Chimera: A unified framework for
  {{B}}/{{FV}}, {{TFHE}} and {{HEAAN}} fully homomorphic encryption and
  predictions for deep learning,'' Aug. 2018. [Online]. Available:
  \url{https://eprint.iacr.org/2018/758}
\BIBentrySTDinterwordspacing

\bibitem{Costache2019-jd}
\BIBentryALTinterwordspacing
A.~Costache, K.~Laine, and R.~Player, ``Evaluating the effectiveness of
  heuristic worst-case noise analysis in {{FHE}},'' 2019. [Online]. Available:
  \url{https://eprint.iacr.org/2019/493}
\BIBentrySTDinterwordspacing

\bibitem{Gentry2012b}
\BIBentryALTinterwordspacing
C.~Gentry, S.~Halevi, and N.~P. Smart, ``Homomorphic {{Evaluation}} of the
  {{AES Circuit}},'' in \emph{Annual Cryptology Conference}.\hskip 1em plus
  0.5em minus 0.4em\relax {Springer}, 2012, pp. 850--867. [Online]. Available:
  \url{https://link.springer.com/chapter/10.1007/978-3-642-32009-5_49}
\BIBentrySTDinterwordspacing

\bibitem{Lauter2014-tv}
\BIBentryALTinterwordspacing
K.~Lauter, A.~L{\'o}pez-Alt, and M.~Naehrig, ``Private computation on encrypted
  genomic data,'' in \emph{Progress in Cryptology - {LATINCRYPT} 2014}.\hskip
  1em plus 0.5em minus 0.4em\relax Springer International Publishing, 2014, pp.
  3--27. [Online]. Available:
  \url{http://dx.doi.org/10.1007/978-3-319-16295-9_1}
\BIBentrySTDinterwordspacing

\bibitem{Halevi2014-cb}
\BIBentryALTinterwordspacing
S.~Halevi and V.~Shoup, ``Algorithms in {HElib},'' in \emph{Advances in
  Cryptology -- {CRYPTO} 2014}.\hskip 1em plus 0.5em minus 0.4em\relax Springer
  Berlin Heidelberg, 2014, pp. 554--571. [Online]. Available:
  \url{http://dx.doi.org/10.1007/978-3-662-44371-2_31}
\BIBentrySTDinterwordspacing

\bibitem{crockett2018alchemy}
E.~Crockett, C.~Peikert, and C.~Sharp, ``{{ALCHEMY}}: {{A Language}} and
  {{Compiler}} for {{Homomorphic Encryption Made easY}},'' in \emph{Proceedings
  of the 2018 {{ACM SIGSAC}} Conference on Computer and Communications
  Security}.\hskip 1em plus 0.5em minus 0.4em\relax {ACM}, 2018, pp.
  1020--1037.

\bibitem{Viand2018-cs}
\BIBentryALTinterwordspacing
A.~Viand and H.~Shafagh, ``Marble: {{Making}} fully homomorphic encryption
  accessible to all,'' in \emph{Proceedings of the 6th Workshop on Encrypted
  Computing \& Applied Homomorphic Cryptography}.\hskip 1em plus 0.5em minus
  0.4em\relax {ACM}, Oct. 2018, pp. 49--60. [Online]. Available:
  \url{https://dl.acm.org/citation.cfm?doid=3267973.3267978}
\BIBentrySTDinterwordspacing

\bibitem{Carpov2015-ok}
\BIBentryALTinterwordspacing
S.~Carpov, P.~Dubrulle, and R.~Sirdey, ``Armadillo: {{A}} compilation chain for
  privacy preserving applications,'' in \emph{Proceedings of the 3rd
  International Workshop on Security in Cloud Computing}, ser. {{SCC}}
  '15.\hskip 1em plus 0.5em minus 0.4em\relax {New York, NY, USA}: {ACM}, 2015,
  pp. 13--19. [Online]. Available:
  \url{http://doi.acm.org/10.1145/2732516.2732520}
\BIBentrySTDinterwordspacing

\bibitem{crockett2016lambdaolambda}
\BIBentryALTinterwordspacing
E.~Crockett and C.~Peikert, ``{{$\Lambda$}}o{{$\lambda$}}: {{Functional}}
  lattice cryptography,'' in \emph{Proceedings of the 2016 {{ACM SIGSAC}}
  Conference on Computer and Communications Security}.\hskip 1em plus 0.5em
  minus 0.4em\relax {ACM}, 2016, pp. 993--1005. [Online]. Available:
  \url{https://dl.acm.org/doi/abs/10.1145/2976749.2978402}
\BIBentrySTDinterwordspacing

\bibitem{Chillotti2020-ll}
\BIBentryALTinterwordspacing
I.~Chillotti, M.~Joye, D.~Ligier, J.-B. Orfila, and S.~Tap, ``{CONCRETE}:
  Concrete operates on ciphertexts rapidly by extending {TfhE},'' in
  \emph{{WAHC} 2020 -- 8th Workshop on Encrypted Computing \& Applied
  Homomorphic Cryptography}, 15~Dec. 2020. [Online]. Available:
  \url{https://homomorphicencryption.org/wp-content/uploads/2020/12/wahc20_demo_damien.pdf}
\BIBentrySTDinterwordspacing

\bibitem{Ducas2015-ei}
\BIBentryALTinterwordspacing
L.~Ducas and D.~Micciancio, ``{FHEW}: Bootstrapping homomorphic encryption in
  less than a second,'' in \emph{Advances in Cryptology -- {EUROCRYPT}
  2015}.\hskip 1em plus 0.5em minus 0.4em\relax Springer Berlin Heidelberg,
  2015, pp. 617--640. [Online]. Available:
  \url{http://dx.doi.org/10.1007/978-3-662-46800-5_24}
\BIBentrySTDinterwordspacing

\bibitem{CryptoExperts2016-yq}
\BIBentryALTinterwordspacing
{CryptoExperts}, ``{FV-NFLlib},'' May 2016. [Online]. Available:
  \url{https://github.com/CryptoExperts/FV-NFLlib}
\BIBentrySTDinterwordspacing

\bibitem{Halevi2014-bh}
\BIBentryALTinterwordspacing
S.~Halevi and V.~Shoup, ``\BIBforeignlanguage{English}{Algorithms in
  {{HElib}}},'' in \emph{\BIBforeignlanguage{English}{Advances in Cryptology
  \textendash{} {{CRYPTO}} 2014}}, ser. Lecture Notes in Computer Science,
  J.~A. Garay and R.~Gennaro, Eds.\hskip 1em plus 0.5em minus 0.4em\relax
  {Springer, Berlin, Heidelberg / Springer}, Aug. 2014, pp. 554--571. [Online].
  Available: \url{http://link.springer.com/chapter/10.1007/978-3-662-44371-
  2_31}
\BIBentrySTDinterwordspacing

\bibitem{Mouchet2020-kh}
\BIBentryALTinterwordspacing
C.~Mouchet and J.-P. Bossuat, ``Lattigo: A multiparty homomorphic encryption
  library in go,'' in \emph{{WAHC} 2020 -- 8th Workshop on Encrypted Computing
  \& Applied Homomorphic Cryptography}, 15~Dec. 2020. [Online]. Available:
  \url{https://homomorphicencryption.org/wp-content/uploads/2020/12/wahc20_demo_christian.pdf}
\BIBentrySTDinterwordspacing

\bibitem{polyakov2019palisade}
\BIBentryALTinterwordspacing
Y.~Polyakov, K.~Rohloff, and G.~W. Ryan, ``{{PALISADE Lattice Cryptography
  Library User Manual}} (v1.6.0),'' Tech. Rep., Sep. 2019. [Online]. Available:
  \url{https://palisade-crypto.org/documentation}
\BIBentrySTDinterwordspacing

\bibitem{Vernam_Group2018-mf}
\BIBentryALTinterwordspacing
{Vernam Group}, ``{cuFHE},'' Mar. 2018. [Online]. Available:
  \url{https://github.com/vernamlab/cuFHE}
\BIBentrySTDinterwordspacing

\bibitem{NuCypher2019-dc}
\BIBentryALTinterwordspacing
{NuCypher}, ``nufhe,'' 19~Jul. 2019. [Online]. Available:
  \url{https://github.com/nucypher/nufhe}
\BIBentrySTDinterwordspacing

\bibitem{Shoup2016}
\BIBentryALTinterwordspacing
V.~Shoup and {Others}, ``{{NTL}}: {{A}} library for doing number theory,'' Aug.
  2016. [Online]. Available: \url{http://www.shoup.net/ntl/}
\BIBentrySTDinterwordspacing

\bibitem{Halevi2015}
\BIBentryALTinterwordspacing
S.~Halevi and V.~Shoup, ``\BIBforeignlanguage{English}{Bootstrapping for
  {{HElib}}},'' in \emph{\BIBforeignlanguage{English}{Advances in Cryptology
  \textendash{} {{EUROCRYPT}} 2015}}, ser. Lecture Notes in Computer Science,
  E.~Oswald and M.~Fischlin, Eds., {Springer}.\hskip 1em plus 0.5em minus
  0.4em\relax {Springer, Berlin, Heidelberg / Springer}, Apr. 2015, pp.
  641--670. [Online]. Available:
  \url{http://link.springer.com/chapter/10.1007/978-3-662-46800-5_25}
\BIBentrySTDinterwordspacing

\bibitem{Halevi2020-ok}
\BIBentryALTinterwordspacing
------, ``{HElib} design principles,'' Tech. Rep., 2020. [Online]. Available:
  \url{https://homenc.github.io/HElib/documentation/Design_Document/HElib-design.pdf}
\BIBentrySTDinterwordspacing

\bibitem{player2018parameter}
\BIBentryALTinterwordspacing
R.~Player, ``Parameter selection in lattice-based cryptography,'' Ph.D.
  dissertation, PhD thesis, Royal Holloway, University of London, 2018.
  [Online]. Available:
  \url{https://pure.royalholloway.ac.uk/portal/files/29983580/2018playerrphd.pdf}
\BIBentrySTDinterwordspacing

\bibitem{chielle2018e3}
\BIBentryALTinterwordspacing
E.~Chielle, O.~Mazonka, N.~G. Tsoutsos, and M.~Maniatakos, ``E3: {{A
  Framework}} for {{Compiling C}}++ {{Programs}} with {{Encrypted Operands}},''
  \emph{IACR Cryptology ePrint Archive}, vol. 2018, p. 1013, 2018. [Online].
  Available: \url{https://eprint.iacr.org/2018/1013}
\BIBentrySTDinterwordspacing

\bibitem{Archer2019-iy}
\BIBentryALTinterwordspacing
D.~W. Archer, J.~M. Calder{\'o}n~Trilla, J.~Dagit, A.~Malozemoff, Y.~Polyakov,
  K.~Rohloff, and G.~Ryan, ``{{RAMPARTS}}: {{A Programmer}}-{{Friendly}} system
  for building homomorphic encryption applications,'' in \emph{Proceedings of
  the 7th {{ACM}} Workshop on Encrypted Computing \& Applied Homomorphic
  Cryptography - {{WAHC}}'19}.\hskip 1em plus 0.5em minus 0.4em\relax {New
  York, New York, USA}: {ACM Press}, 2019, pp. 57--68. [Online]. Available:
  \url{http://dl.acm.org/citation.cfm?doid=3338469.3358945}
\BIBentrySTDinterwordspacing

\bibitem{dathathri2018chet}
\BIBentryALTinterwordspacing
R.~Dathathri, O.~Saarikivi, H.~Chen, K.~Laine, K.~Lauter, S.~Maleki,
  M.~Musuvathi, and T.~Mytkowicz, ``{CHET}: an optimizing compiler for
  fully-homomorphic neural-network inferencing,'' in \emph{Proceedings of the
  40th {ACM} {SIGPLAN} Conference on Programming Language Design and
  Implementation}.\hskip 1em plus 0.5em minus 0.4em\relax New York, NY, USA:
  ACM, 8~Jun. 2019, pp. 142--156. [Online]. Available:
  \url{https://dl.acm.org/citation.cfm?doid=3314221.3314628}
\BIBentrySTDinterwordspacing

\bibitem{van2019sealion}
\BIBentryALTinterwordspacing
T.~{van Elsloo}, G.~Patrini, and H.~{Ivey-Law}, ``{{SEALion}}: {{A Framework}}
  for {{Neural Network Inference}} on {{Encrypted Data}},'' \emph{arXiv
  preprint arXiv:1904.12840}, 2019. [Online]. Available:
  \url{https://arxiv.org/abs/1904.12840}
\BIBentrySTDinterwordspacing

\bibitem{Crockett2017}
\BIBentryALTinterwordspacing
E.~Crockett, ``Simply safe lattice cryptography,'' Ph.D. dissertation, Georgia
  Institute of Technology, 2017. [Online]. Available:
  \url{https://smartech.gatech.edu/handle/1853/58734}
\BIBentrySTDinterwordspacing

\bibitem{Herbert2019-ci}
\BIBentryALTinterwordspacing
V.~Herbert, ``Automatize parameter tuning in
  {{Ring}}-{{Learning}}-{{With}}-{{Errors}}-based leveled homomorphic
  cryptosystem implementations,'' 2019. [Online]. Available:
  \url{https://eprint.iacr.org/2019/1402}
\BIBentrySTDinterwordspacing

\bibitem{mishchenko2018abc}
\BIBentryALTinterwordspacing
A.~Mishchenko, ``{{ABC}}: {{System}} for sequential logic synthesis and formal
  verification,'' 2018. [Online]. Available:
  \url{https://github.com/berkeley-abc/abc}
\BIBentrySTDinterwordspacing

\bibitem{carpov2017multi}
\BIBentryALTinterwordspacing
S.~Carpov, P.~Aubry, and R.~Sirdey, ``A multi-start heuristic for
  multiplicative depth minimization of boolean circuits,'' in
  \emph{International Workshop on Combinatorial Algorithms}.\hskip 1em plus
  0.5em minus 0.4em\relax {Springer}, 2017, pp. 275--286. [Online]. Available:
  \url{https://link.springer.com/chapter/10.1007/978-3-319-78825-8_23}
\BIBentrySTDinterwordspacing

\bibitem{Aubry2020-jy}
\BIBentryALTinterwordspacing
P.~Aubry, S.~Carpov, and R.~Sirdey, ``Faster homomorphic encryption is not
  enough: {{Improved}} heuristic for multiplicative depth minimization of
  boolean circuits,'' in \emph{Topics in Cryptology \textendash{}
  {{CT}}-{{RSA}} 2020}.\hskip 1em plus 0.5em minus 0.4em\relax {Springer
  International Publishing}, 2020, pp. 345--363. [Online]. Available:
  \url{http://dx.doi.org/10.1007/978-3-030-40186-3_15}
\BIBentrySTDinterwordspacing

\bibitem{Lee2020-uq}
\BIBentryALTinterwordspacing
D.~Lee, W.~Lee, H.~Oh, and K.~Yi, ``Optimizing homomorphic evaluation circuits
  by program synthesis and term rewriting,'' in \emph{Proceedings of the 41st
  {{ACM SIGPLAN}} Conference on Programming Language Design and
  Implementation}, ser. {{PLDI}} 2020.\hskip 1em plus 0.5em minus 0.4em\relax
  {New York, NY, USA}: {Association for Computing Machinery}, Nov. 2020, pp.
  503--518. [Online]. Available: \url{https://doi.org/10.1145/3385412.3385996}
\BIBentrySTDinterwordspacing

\bibitem{boemer2019}
\BIBentryALTinterwordspacing
F.~Boemer, Y.~Lao, R.~Cammarota, and C.~Wierzynski, ``{{nGraph}}-{{HE}}: {{A
  Graph Compiler}} for {{Deep Learning}} on {{Homomorphically Encrypted
  Data}},'' in \emph{Proceedings of the 16th {{ACM}} International Conference
  on Computing Frontiers}, ser. {{CF}} '19.\hskip 1em plus 0.5em minus
  0.4em\relax {New York, NY, USA}: {ACM}, 2019, pp. 3--13. [Online]. Available:
  \url{https://dl.acm.org/doi/10.1145/3310273.3323047}
\BIBentrySTDinterwordspacing

\bibitem{cyphers2018intel}
\BIBentryALTinterwordspacing
S.~Cyphers, A.~K. Bansal, A.~Bhiwandiwalla, J.~Bobba, M.~Brookhart,
  A.~Chakraborty, W.~Constable, C.~Convey, L.~Cook, O.~Kanawi \emph{et~al.},
  ``Intel {{nGraph}}: {{An}} intermediate representation, compiler, and
  executor for deep learning,'' \emph{arXiv preprint arXiv:1801.08058}, 2018.
  [Online]. Available: \url{https://arxiv.org/abs/1801.08058}
\BIBentrySTDinterwordspacing

\bibitem{Carpov2016-tx}
\BIBentryALTinterwordspacing
S.~Carpov, T.~H. Nguyen, R.~Sirdey, G.~Constantino, and F.~Martinelli,
  ``Practical {{Privacy}}-{{Preserving}} medical diagnosis using homomorphic
  encryption,'' in \emph{2016 {{IEEE}} 9th International Conference on Cloud
  Computing ({{CLOUD}})}, Jun. 2016, pp. 593--599. [Online]. Available:
  \url{http://dx.doi.org/10.1109/CLOUD.2016.0084}
\BIBentrySTDinterwordspacing

\bibitem{lecun-mnisthandwrittendigit-2010}
\BIBentryALTinterwordspacing
Y.~LeCun and C.~Cortes, ``{{MNIST}} handwritten digit database,'' 2010.
  [Online]. Available: \url{http://yann.lecun.com/exdb/mnist/}
\BIBentrySTDinterwordspacing

\bibitem{Gilad-Bachrach2016-aq}
\BIBentryALTinterwordspacing
R.~{Gilad-Bachrach}, N.~Dowlin, K.~Laine, K.~Lauter, M.~Naehrig, and
  J.~Wernsing, ``\BIBforeignlanguage{English}{{{CryptoNets}}: {{Applying}}
  neural networks to encrypted data with high throughput and accuracy},'' in
  \emph{\BIBforeignlanguage{English}{Proceedings of the 33rd International
  Conference on Machine Learning}}, vol.~48.\hskip 1em plus 0.5em minus
  0.4em\relax {New York, New York, USA}: {PMLR}, 2016, pp. 201--210. [Online].
  Available: \url{http://proceedings.mlr.press/v48/gilad-bachrach16.html}
\BIBentrySTDinterwordspacing

\bibitem{lecun1998gradient}
\BIBentryALTinterwordspacing
Y.~LeCun, L.~Bottou, Y.~Bengio, and P.~Haffner, ``Gradient-based learning
  applied to document recognition,'' \emph{Proceedings of the IEEE}, vol.~86,
  no.~11, pp. 2278--2324, 1998. [Online]. Available:
  \url{https://ieeexplore.ieee.org/document/726791}
\BIBentrySTDinterwordspacing

\bibitem{savage1997models}
J.~E. Savage, \emph{Models of Computation: {{Exploring}} the Power of
  Computing}, 1st~ed.\hskip 1em plus 0.5em minus 0.4em\relax {Boston, MA, USA}:
  {Addison-Wesley Longman Publishing Co., Inc.}, 1997.

\bibitem{Chillotti2018-ds}
\BIBentryALTinterwordspacing
I.~Chillotti, N.~Gama, M.~Georgieva, and M.~Izabach{\`e}ne, ``{{TFHE}}:
  {{Fast}} fully homomorphic encryption over the torus,'' 2018. [Online].
  Available: \url{https://eprint.iacr.org/2018/421}
\BIBentrySTDinterwordspacing

\bibitem{Cheon2020-remarks}
J.~H. Cheon, S.~Hong, and D.~Kim, ``Remark on the security of ckks scheme in
  practice,'' Cryptology ePrint Archive, Report 2020/1581, 2020,
  \url{https://eprint.iacr.org/2020/1581}.

\bibitem{Li2020}
B.~Li and D.~Micciancio, ``On the security of homomorphic encryption on
  approximate numbers,'' Cryptology ePrint Archive, Report 2020/1533, 2020,
  \url{https://eprint.iacr.org/2020/1533}.

\bibitem{Cheon2019-if}
\BIBentryALTinterwordspacing
J.~H. Cheon, K.~Han, A.~Kim, M.~Kim, and Y.~Song, ``A full {{RNS}} variant of
  approximate homomorphic encryption,'' in \emph{Selected Areas in Cryptography
  \textendash{} {{SAC}} 2018}.\hskip 1em plus 0.5em minus 0.4em\relax {Springer
  International Publishing}, 2019, pp. 347--368. [Online]. Available:
  \url{http://dx.doi.org/10.1007/978-3-030-10970-7_16}
\BIBentrySTDinterwordspacing

\end{thebibliography}
\appendices
\pagebreak
\section{}
\label{fhe-appendix}
In this appendix, we briefly introduce the notion of FHE and outline three important modern schemes.
We focus primarily on aspects relevant to FHE application developers, i.e., plaintext spaces, encodings, and aspects that impact performance.

\subsection{Fully Homomorphic Encryption}

A \emph{homomorphic} encryption scheme is a (most frequently public-key) encryption scheme where there exists a homomorphism between operations on the plaintext and operations on the ciphertext:
\begin{equation*}
    \Dec(\Enc(x + y)) = \Dec(\Enc(x) \oplus \Enc(y))
\end{equation*}
where $+$ and $\oplus$ are operations over the plaintext and ciphertext space, respectively.
A \emph{fully homomorphic} encryption scheme is one that is homomorphic in regards to both addition and multiplication.
We omit a more formal treatment here and instead refer to~\cite{Gentry2009-zi} for a formal definition, including several constraints that apply to exclude trivial constructions.

Addition and multiplication allow us to compute any polynomial function over the encrypted data 
but many frequently-used functions like comparisons or sorting are non-polynomial, i.e., cannot (easily) be expressed as polynomial functions.
However, multiplication and addition in $\mathbb{Z}_2$ can be used to emulate \texttt{AND}- and \texttt{XOR}-gates, respectively.
Together with memory, this is Turing-complete, i.e., one can emulate arbitrary computations~\cite{savage1997models}.

\subsection{FHE Schemes}

We briefly introduce three of the most widely used homomorphic encryption schemes.

\subsubsection{CGGI}
The \acl{cggi} scheme~\cite{Chillotti2016Dec,Chillotti2017-yd} is part of a third generation of \ac{fhe} schemes based on the \acf{gsw} scheme~\cite{gentry2013homomorphic}. \acused{cggi}
More commonly known as TFHE, we refer to it here by the author initials in order to avoid confusion with the TFHE library.

In \ac{cggi}, the plaintext and ciphertext space $T$ is a group of polynomials (modulo some irreducible polynomial) of degree up to $n-1$ over the torus $\T = \R /  \Z$ (i.e., the real numbers $\mod 1$).
The message space is generally chosen so that the computation emulates binary circuits and homomorphic addition becomes \Xor and multiplication becomes \AndGate. 
Since $T$ is not a ring, it supports addition but has no native multiplication operation.
However,
 multiplications are defined between \ac{gsw} ciphertexts and ciphertexts in $T$.
This is used to perform multiplications and non-linear operations over ciphertexts in $T$ during the bootstrapping process, 
by encrypting the bootstrapping key as a \ac{gsw} ciphertext.
Multiplications between ciphertexts in $T$ are realized as one specific type of such a non-linear transformation applied during bootstrapping.
In this \emph{gate-bootstrapped} version of the scheme, every non-linear gate therefore inherently includes bootstrapping.

Chillotti et al. also show how to construct a MUX gate that selects between two ciphertexts in $T$ dependent on a \ac{gsw} ciphertext and introduce efficient designs for Look-Up-Tables (LUTs).
Finally, they show how to use weighted Finite Automata to emulate binary multiplication~\cite{Chillotti2018-ds}.
However, these techniques are not implemented in the \ac{tfhe} library.

\subsubsection{BFV}
The \acl{bfv} \acused{bfv} scheme is a second-generation scheme.
Fan and Vercauteren~\cite{fan2012somewhat} ported a scheme by Brakerski~\cite{brakerski2012fully} to the ring-LWE domain and improved its performance.
In \ac{bfv}, the plaintext space $R_t$ is a ring of polynomials (modulo some irreducible polynomial) of degree up to $n-1$ with coefficients in $\Z_t$.
Note that for $t=2$, we are in the binary circuit setting.
Messages $m \in \Z_t$ can be encoded into this plaintext space as a constant polynomial $f(x) = m$.
However, this is inefficient as only one of $n$ coefficients is utilized.
Simply encoding messages into additional coefficients raises issues when performing computations:
while polynomial additions work coefficient-wise, multiplications combine different coefficients in undesired ways.
Instead, one can achieve SIMD-style \emph{batching} via the Chinese Remainder Theorem~\cite{iliashenko2019optimisations}.
By choosing $n = \Pi_{i=0}^k n_i$ , a degree-$n$ polynomial can be reinterpreted as the multiplication of $k$ lower-degree polynomials.
Using this technique, $k$ messages can be packed into a single plaintext, where $k \gg 1000$ in practice,  while maintaining meaningful semantics.
Automorphisms additionally enable homomorphic rotations of the elements~\cite{halevi2018faster}.

The ciphertexts, meanwhile, are made up of at elements from $R_q$, which has the same structure as $R_p$, but with a different coefficient modulus $q$.
Each ciphertexts consists of at least two elements, i.e., $c = [c_0, c_1]$.
These polynomials $c_i$ can themselves be interpreted as coefficients of a polynomial $C(X)$.
Homomorphic addition and multiplication between ciphertexts correspond to addition and multiplication between the $C(X)$'s, respectively.
As a consequence, the result of a multiplication is a quadratic polynomial, i.e., a ciphertext with three elements $c = [c_0, c_1,c_2]$.
During further multiplications the noise term would first become squared, then cubed, etc. growing excessively.
Therefore, BFV and similar schemes introduce a \emph{relinearization} procedure to transform ciphertexts back to linear form.
We omit a description of bootstrapping and instead note that BFV is more commonly used in leveled mode where the parameters are chosen sufficiently large to complete the computation without bootstrapping.

\subsubsection{CKKS}
The \acl{ckks} scheme~\cite{cheon2017homomorphic}, also known as \acf{heaan}, focuses on homomorphic encryption for \emph{approximate} numbers. \acused{ckks}
Formally speaking it is not an FHE scheme since it only fulfills the requirements approximately, i.e., 
$\Dec(\Enc(x + y)) \approx \Dec(\Enc(x) \oplus \Enc(y))$, for some operations $+$ and $\oplus$.
While this slight relaxation has led to an extremely efficient scheme, some care must be taken when using approximate FHE schemes~\cite{Cheon2020-remarks,Li2020}.
\ac{ckks} is designed primarily for computations with fixed point numbers, i.e., a number $x$ is represented as $m = \nearest{x*\varDelta}$ for \emph{scale} $\varDelta$, usually a large integer.
While any integer-based scheme can be used for fixed-point computations, they quickly run into overflow issues.
\ac{ckks} addresses this by introducing a homomorphic rounding operation that reduces the scale of a product back to the original scale $\varDelta$.

In \ac{ckks}, the logical message space is $\C^{n}$, i.e., vectors over the complex numbers, although most applications use only the real part.
The plaintext space $R$  is a ring of polynomials (modulo some irreducible polynomial) of degree up to $n-1$ with coefficients in $\Z$.
Given a scaling factor $\varDelta \in \R$, we represent  $m \in \R$ as $m' = \nearest{\varDelta m} \in \Z$.
For brevity, we skip a description of the encoding of such representations into a plaintext polynomial and simply note that the encoding introduces small additional approximation errors.
During encryption, noise is intentionally introduced, but this noise overlaps with the least significant bits of the plaintext.
Therefore, the approximation error and noise are treated as one, and rather than suddenly losing the message when the noise reaches a threshold, we gradually lose accuracy.

Like in \ac{bfv}, ciphertexts in \ac{ckks} are arrays of elements $c_i \in R_q$ and multiplications require relinearization.
However, different to \ac{bfv}, the noise  $e$ grows quadratically with each subsequent multiplication.
After $\ell$ multiplications, it has grown to $e ^{2^{\ell}}$ and a modulus $q \approx e^{2^{\ell}}$ would be required to decrypt the resulting ciphertext correctly.
Instead, one can scale the ciphertext down by a factor $\omega$, i.e., go from $R_q$ to $R_{q/\omega}$.
This is known as \emph{rescaling} and is similar to the \emph{modulus switching} operation in the \acf{bgv} scheme~\cite{Brakerski2014-uq} but rescaling also affects the plaintext.
Using rescaling, a modulus of size $(\ell+1)\omega e$ suffices to evaluate $\ell$ subsequent multiplications.
During this operation, the plaintext encrypted in the ciphertext is also effectively rescaled to $\varDelta' = \nearest{\varDelta/\omega}$.
Choosing $q = \Pi_{i=0}^k q_i$ where $q_i$ are roughly equally sized primes improves both performance~\cite{Cheon2019-if} and, by setting $\omega = q_i$, ensures a (nearly) constant scale throughout the computation.

\end{document}